\documentclass[twocolumn,useAMS,usenatbib]{mn2e}

\usepackage{natbib}

\usepackage{fmtcount}
\usepackage{threeparttable}
\usepackage{graphicx,subfigure}
\usepackage{multirow}
\usepackage{url}

\title[SN 2009N in NGC 4487]{SN 2009N: Linking normal and subluminous type II-P SNe}

\author[K. Tak\' ats et al.]{K. Tak\' ats$^{1,2}$\thanks{E-mail: ktakats@gmail.com}, M. L. Pumo$^{3,4}$,  N. Elias-Rosa$^{5,3}$, A. Pastorello$^3$, G. Pignata$^{1}$,\newauthor E. Paillas$^1$, L. Zampieri$^3$, J. P. Anderson$^6$, J. Vink\'o$^{2}$, S. Benetti$^3$, \newauthor M-T. Botticella$^{7,8}$,  F. Bufano$^{1,3}$, A. Campillay$^9$, R. Cartier$^6$,  M. Ergon$^{10}$,\newauthor G. Folatelli$^{11}$, R. J. Foley$^{12,13}$, F. F\" orster$^6$, M. Hamuy$^6$, V-P. Hentunen$^{14}$, \newauthor E. Kankare$^{15}$, G. Leloudas$^{16,17}$, N. Morrell$^9$, M. Nissinen$^{14}$, M. M. Phillips$^{9}$,\newauthor S.~J. Smartt$^8$, M. Stritzinger$^{18}$,  S. Taubenberger$^{19}$, S. Valenti$^{20,21}$,  S. D. Van Dyk$^{22}$, \newauthor J.~B. Haislip$^{23}$, A. P. LaCluyze$^{23}$, J.~P. Moore$^{23}$, D. Reichart$^{23}$\\
$^1$Departamento de Ciencias Fisicas, Universidad Andres Bello, Avda. Republica 252, Santiago, Chile \\
$^2$Department of Optics $\&$ Quantum Electronics, University of Szeged, D\'om t\'er 9., Szeged, 6720 Hungary\\
$^3$INAF - Osservatorio Astronomico di Padova, vicolo dell'Osservatorio 5, 35122 Padova, Italy\\
$^4$INAF - Osservatorio Astrofisico di Catania, via S. Sofia 78, 95123 Catania, Italy\\
$^5$Institut de Ci\`encies de l'Espai (CSIC-IEEC), Facultat de Ci\`encies, Campus UAB, 08193 Bellaterra, Spain\\
$^6$Departamento de Astronom\'ia, Universidad de Chile, Casilla 36-D, Santiago, Chile\\
$^7$INAF - Osservatorio Astronomico di Capodimonte, Salita Moiariello, 80128 Napoli, Italy\\
$^8$Astrophysics Research Centre, School of Mathematics and Physics, Queen's University Belfast, Belfast BT7 1NN, UK\\
$^9$Las Campanas Observatory, Carnegie Observatories, Casilla 601, La Serena, Chile\\
$^{10}$The Oskar Klein Centre, Department of Astronomy, AlbaNova, Stockholm University, 10691 Stockholm, Sweden\\
$^{11}$Kavli Institute for the Physics and Mathematics of the Universe, Todai Institutes for Advanced Study (TODIAS), University of Tokyo,\\
5-1-5 Kashiwanoha, Kashiwa, Chiba 277-8583, Japan\\
$^{12}$Astronomy Department, University of Illinois at Urbana-Champaign, 1002 W.\ Green Street, Urbana, IL 61801 USA\\
$^{13}$Department of Physics, University of Illinois at Urbana-Champaign, 1110 W.\ Green Street, Urbana, IL 61801 USA\\
$^{14}$Taurus Hill Observatory, H\"ark\"am\"aentie 88, 79480 Kangaslampi, Finland\\
$^{15}$Finnish Centre for Astronomy with ESO (FINCA), University of Turku, V\"ais\"al\"antie 20, FI-21500 Piikki\"o, Finland\\
$^{16}$The Oskar Klein Centre, Department of Physics, Stockholm University, 106 91 Stockholm, Sweden\\ 
$^{17}$Dark Cosmology Centre, Niels Bohr Institute, University of Copenhagen, Juliane Maries vej 30, 2100 Copenhagen, Denmark\\
$^{18}$Department of Physics and Astronomy, Aarhus University, Ny Munkegade 120, DK-8000 Aarhus C, Denmark\\
$^{19}$Max Planck Institut f\"ur Astrophysik, Karl-Schwarzschild-Str. 1, 85741 Garching bei M\"unchen, Germany\\
$^{20}$Las Cumbres Observatory Global Telescope Network, 6740 Cortona Dr., Suite 102, Goleta, CA 93117, USA\\
$^{21}$Department of Physics, University of California, Santa Barbara, Broida Hall, Mail Code 9530, Santa Barbara, CA 93106-9530, USA\\
$^{22}$Spitzer Science Center, California Institute of Technology, 1200 East California Boulevard, Pasadena, CA 91125, USA\\
$^{23}$University of North Carolina at Chapel Hill, Campus Box 3255, Chapel Hill, NC 27599-3255, USA
}

\begin{document}

\date{Accepted  Received ; in original form }

\pagerange{\pageref{firstpage}--\pageref{lastpage}} \pubyear{}

\maketitle

\label{firstpage}

\begin{abstract}

We present ultraviolet, optical, near-infrared photometry and spectroscopy of SN~2009N in NGC~4487. This object is a type II-P supernova with spectra resembling those of subluminous II-P supernovae, while its  bolometric luminosity is similar to that of the intermediate luminosity SN~2008in. We created {\sc synow} models of the plateau phase spectra for line identification and to measure the expansion velocity. 
In the near-infrared spectra we find signs indicating possible weak interaction between the supernova ejecta and the pre-existing circumstellar material. These signs are also present in the previously unpublished near-infrared spectra of SN~2008in. 
The distance to SN~2009N is determined via the expanding photosphere method and the standard candle method as $D= 21.6 \pm 1.1$~Mpc. The produced nickel-mass is estimated to be $\sim 0.020 \pm 0.004$~M$_{\sun}$. We infer the physical properties of the progenitor at the explosion through hydrodynamical modelling of the observables. We find the values of the total energy as  $\sim 0.48 \times 10^{51}$~erg, the ejected mass as $\sim 11.5$~M$_{\sun}$, and the initial radius as $\sim 287$~R$_{\sun}$.

\end{abstract}


\section{Introduction}\label{introduction}

Type II supernovae (SNe) are classified on the basis of the presence of hydrogen in their spectra. SNe II-P \citep*{Barbon1979} are by far the most frequent, representing about 70 per cent of all SNe type II \citep{li}. They are characterized by nearly constant luminosities during the first period of their evolution (``plateau''). This phase lasts until their thick, initially ionized hydrogen envelopes recombine.

SNe II-P are thought to emerge from stars with a zero age main sequence mass in the range of $8-21$ M$_{\sun}$ \citep{walmswell_eldridge}. In some cases the progenitor star was directly identified in pre-explosion images as a red supergiant (RSG) star \citep{smartt_rev}.
The observational properties of these SNe, such as the peak luminosity, the plateau duration or the expansion velocity display a wide range of values. 

A number of subluminous type II-P events have also been discovered and studied, including SNe 1997D \citep{turatto_ba, benetti_1997D}, 2003Z \citep*{utrobin03Z}, 2005cs \citep{pastorello05csII, takats2006}, 1999br \citep{hamuy_thesis, pastorello99br}, and 2009md \citep{fraser09md}. These SNe have fainter absolute magnitudes, lower expansion velocities, and lower nickel-masses than the majority of SNe II-P. The nature of their progenitors is still debated. 
In the case of SN~1997D, by modelling the observables, \citet{turatto_ba} favored a scenario where the low $^{56}$Ni-mass observed is a result of a fallback of material onto the remnant of the explosion of a massive ($25-40$~M$_{\sun}$) star. 
The hydrodynamical models of \citet{zampieri_sne_conf} inferred the ejecta masses of SNe~1997D and 1999br to be $14$ and $10\,{\rm M_{\sun}}$, respectively, while examining archival pre-explosion images \citet{maund_progs} estimated the mass limit for the progenitor of SN~1999br as $< 15$~M$_{\sun}$. In the case of SN~2003Z, the ejecta mass resulted by the modelling of \citet{utrobin03Z} was $14 \pm 1.2\,{\rm M_{\sun}}$, while \citet{zampieri_sne_conf} estimated it to be $\sim 22\,{\rm M_{\sun}}$.
On the other hand, the progenitors of SNe~2005cs and 2009md were identified in archival images, and were found to have masses of $\sim 8-9$~M$_{\sun}$ \citep{fraser09md}. 

In the last couple of years studies of objects that fit in between normal and subluminous SNe II-P have been published. In particular, SN~2008in had spectra very similar to those of the subluminous SNe II-P, but it had somewhat higher luminosity \citep{roy08in}. The expansion velocities and the amount of the ejected nickel-mass were also between the typical values of normal and subluminous events. Applying the analytical relations of \citet{litvinova}, \citet{roy08in} estimated the mass of the progenitor as $< 20$~M$_{\sun}$, while employing hydrodynamical modelling, \citet{utrobin_08in} obtained the value of $15.5 \pm 2.2\,\rm{M_{\sun}}$.
\citet{gandhi_09js} found that SN~2009js shared the characteristics of both SN~2008in and the subluminous SN~2005cs: the luminosity and the duration of its plateau were more similar to those of SN~2008in, but other properties, i.e. the ejected $^{56}$Ni-mass and the explosion energy were closer to those of SN~2005cs. The mass of the progenitor was estimated in \citet{gandhi_09js} using the relations of \citet{litvinova} as $11 \pm 5$~M$_{\sun}$. 

In this paper we present another ``intermediate-luminosity'' object, SN~2009N. It was discovered by Itagaki on images taken on Jan. 24.86 and 25.62 UT in NGC 4487 \citep{felfed}. They also reported that no source was visible on images taken on 2009 Jan. 3 (limiting magnitude 18). \citet{09N_classif} obtained a spectrum of this SN on Jan. 25, and classified it as a type II SN, adding that the spectrum was a good match to that of SN~2005cs at two days after maximum.

The paper is organized as follows. In Sec. \ref{photometry} we present the photometric data taken with ground-based telescopes through optical and near-infrared filters, and with {\it Swift/UVOT} in the ultraviolet. In Sec. \ref{spectroscopy} we show the optical and near infrared spectroscopic observations and study the spectral evolution.
Based on the observed and measured data, the distance to SN 2009N is calculated in Sec. \ref{distance} via both the expanding photosphere method \citep[EPM,][]{epm_ref} and the standardized candle method \citep[SCM,][]{hamuy_scm}. With the EPM we also estimate the explosion epoch as $t_0 = 2454848.1$~JD (Jan. 16.6 UT), which is adopted throughout the paper.  In Sec. \ref{phys} the main physical parameters of the SN are inferred by hydrodynamical models.
In Sec. \ref{summary} we summarise our results.

%##################################    Photometry   ################################

\section{Photometry}\label{photometry}

Optical photometric data were collected with multiple telescopes using $BVRI$ and $g'r'i'z'$ filters, between days +11 and +413 after explosion (Tables \ref{lc_table}, \ref{lc_table_griz}).

\begin{figure}
\includegraphics[width=84mm]{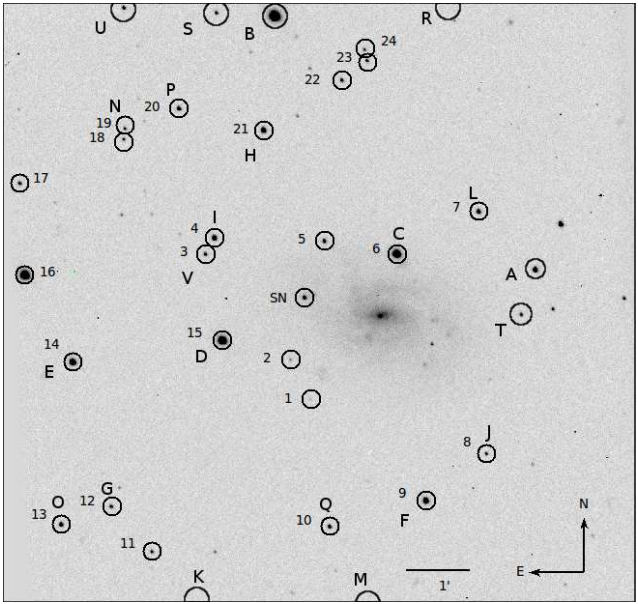}
\caption{The field of SN 2009N in NGC 4478 obtained with one of the PROMPT telescopes in $R$ band. The local comparison stars used to calibrate the optical and NIR photometry are marked with numbers and letters, respectively.}
\label{std_kep}
\end{figure}

The basic data reductions (bias-subtraction, overscan-correction, flat-fielding) were carried out using standard {\sc iraf}\footnote{{\sc iraf} is distributed by the National Optical Astronomy Observatories, which are operated by the Association of Universities for Research in Astronomy, Inc., under the cooperative agreement with the National Science Foundation.} routines. The brightness of the SN was measured using the point-spread function fitting technique.
The calibration of the photometry was performed using standard fields \citep{landolt,landolt2007,sloan_std_fields} observed on photometric nights. Using these images, magnitudes for a local sequence of stars (Fig. \ref{std_kep}, Tables \ref{std} and \ref{std_sloan}) on the SN field were determined and  used to calibrate the SN measurements. 
The data taken with the Liverpool Telescope were reduced using the same process, but with the {\sc quba} pipeline, an {\sc iraf} based Python package \citep[see][for details]{quba}.

Near-infrared (NIR) $YJH$ photometry was obtained with the Swope (RetroCam) and the du-Pont
(WIRC) telescopes at Las Campanas Observatory 
as part of the Carnegie Supernova Project \citep[CSP,][]{hamuy_CSP}, between days $+$10 and $+$427 after explosion. Full details of the survey characteristics, data
reductions, and photometric processing techniques can be found in \citet{hamuy_CSP}, \citet{contreras_CSP}, and \citet{stritzinger_CSP}. Summarising, NIR images were processed through a sequence
of: dark subtraction, flat-field correction, sky-subtraction, non-linearity
correction, then alignment and combination of dithered frames. Photometric
calibration was achieved through observations of standard star fields \citep{persson_nirstandards} and determining the magnitudes of stars on the SN field (Fig~\ref{std_kep}, Table~\ref{std_nir}). The SN magnitudes are reported in Table \ref{nirlc_table}.

\begin{figure*}
\includegraphics[width=\textwidth]{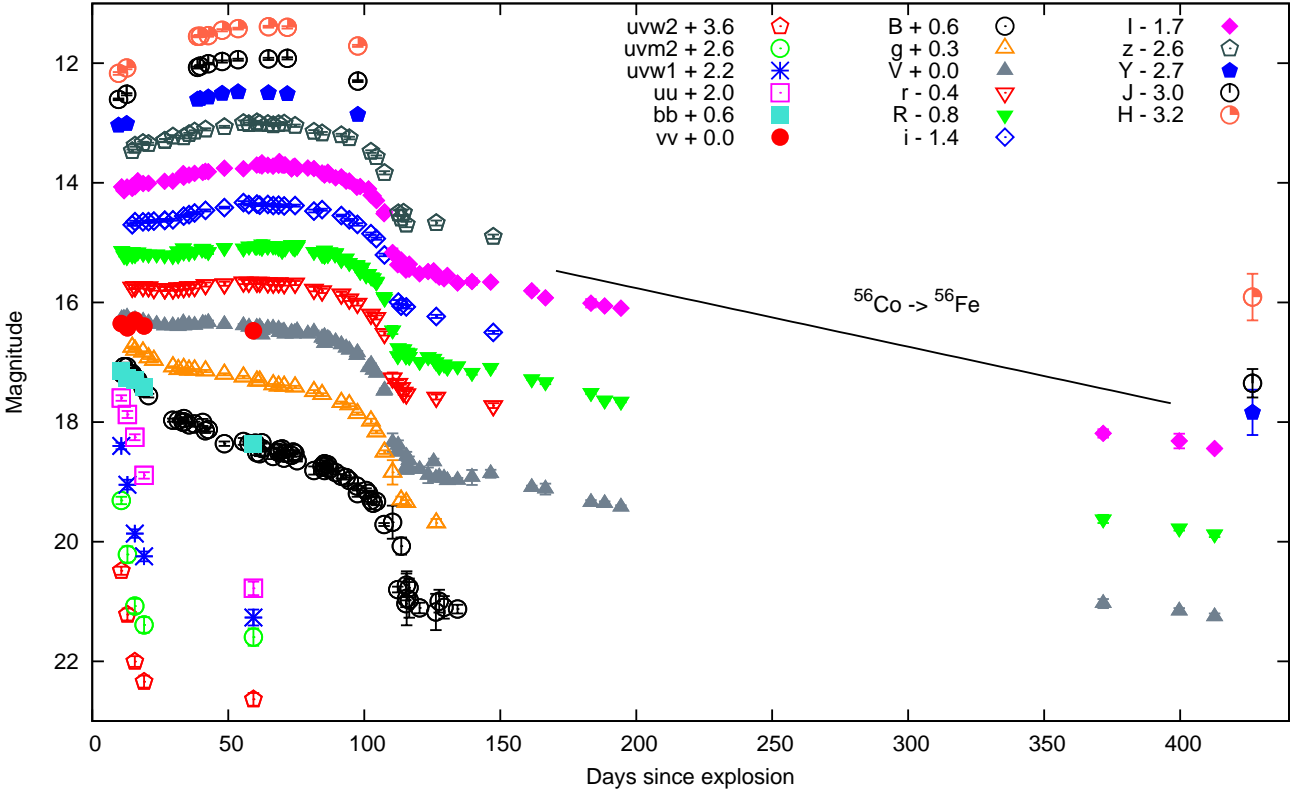}
\caption{UV, optical, NIR light curves of SN 2009N. The explosion epoch is $t_0=2454848.1$~JD.}
\label{lc}
\end{figure*}

The light curve (Fig. \ref{lc}) shows a regular type II-P SN, with a plateau of nearly constant luminosity in $Vv'Rr'$, lasting until about $+110$~days after explosion. The $B$ and $g'$ band light curves show constant decline from early phases, with higher decline rate during the first $\sim 20$ days. In $Ii'YJH$ bands the brightness increases slightly but continuously until about day +70, 
 when it starts to decrease. At the end of the plateau the brightness drops about $2$ magnitudes in $\sim3$ weeks.
The tail phase of SNe II-P is powered by the energy input from the radioactive decay of $^{56}$Co to $^{56}$Fe, so the expected decline rate is 0.98 mag/100~d for complete $\gamma$-ray trapping \citep{patat_1994}. Between days 113-414 we measure the decline rate as $0.85 \pm 0.02$~mag/100~d in the $V$, $1.02 \pm 0.01$~mag/100~d in the $R$, $1.03 \pm 0.02$~mag/100~d in the $I$ band, and $1.02 \pm 0.05$~mag/100~d for the bolometric light curve (Sec.~\ref{sec_bollum}).

\subsection{{\it Swift} photometry}\label{sec_swift}

SN 2009N was also observed with {\it Swift/UVOT} \citep{roming_swift} at 5 epochs with all 6 filters. We used the {\sc heassoft}\footnote{\url{http://heasarc.gsfc.nasa.gov/docs/software/heasoft/}} software to co-add the images and perform aperture photometry. The count rate of the source was measured using a 3$"$ aperture, while the coincidence loss correction was calculated using a 5$"$ aperture. Since there are no template observations to measure the count rate originated from the galaxy, we used an aperture close to the source to measure the background level. An aperture correction was applied from the 3$"$ to the 5$"$ aperture, based on the average PSF available in {\it Swift} CALDB. The zeropoints from \citet{swift_calib} were used to convert the resulted count rate to the UVOT photometric system. The resulted magnitudes are in Table \ref{swift_lc} and shown in Fig. \ref{lc}.

The UV light curve decreases quickly at early times, which is typical of type II-P SNe \citep{brown_swift, dessart2008}, while the $vv$ and $bb$ light curves show similar behaviour than in the Bessel $B$ and $V$ filters.

\subsection{Reddening and Colour Curves}\label{reddening}

The Galactic reddening in the direction of SN 2009N is low, $E(B-V)_{\mathrm{MW}}=0.019 \pm 0.001$~mag \citep{reddening_recalib}. The host galaxy component of the extinction can be estimated from the equivalent width of the Na\,{\sc i} D doublet. 
We have a medium-resolution spectrum, taken on day $+159$ after explosion (see Sec.\ref{spectroscopy}), where the Na\,{\sc i} D$_1$ and D$_2$ lines are resolved. We measured the equivalent widths of these lines as $EW(D1)=0.39 \pm 0.04$~\AA~and $EW(D2)=0.34 \pm 0.03$~\AA. Via the relations determined by \citet{poznanski2012} we calculated the host galaxy extinction as $E(B-V)_{\mathrm host}=0.113 \pm 0.019$~mag. Together with the Galactic component, we adopt the extinction $E(B-V)_{\mathrm tot}=0.13 \pm 0.02$~mag for SN 2009N. We must consider, however, that there is a large scatter for SNe in the EW(NaI) versus E(B-V) plane \citep[e.g.][]{turatto2003, poznanski2011}, therefore the estimation of the extinction is quite uncertain.

\begin{figure}
\includegraphics[width=84mm]{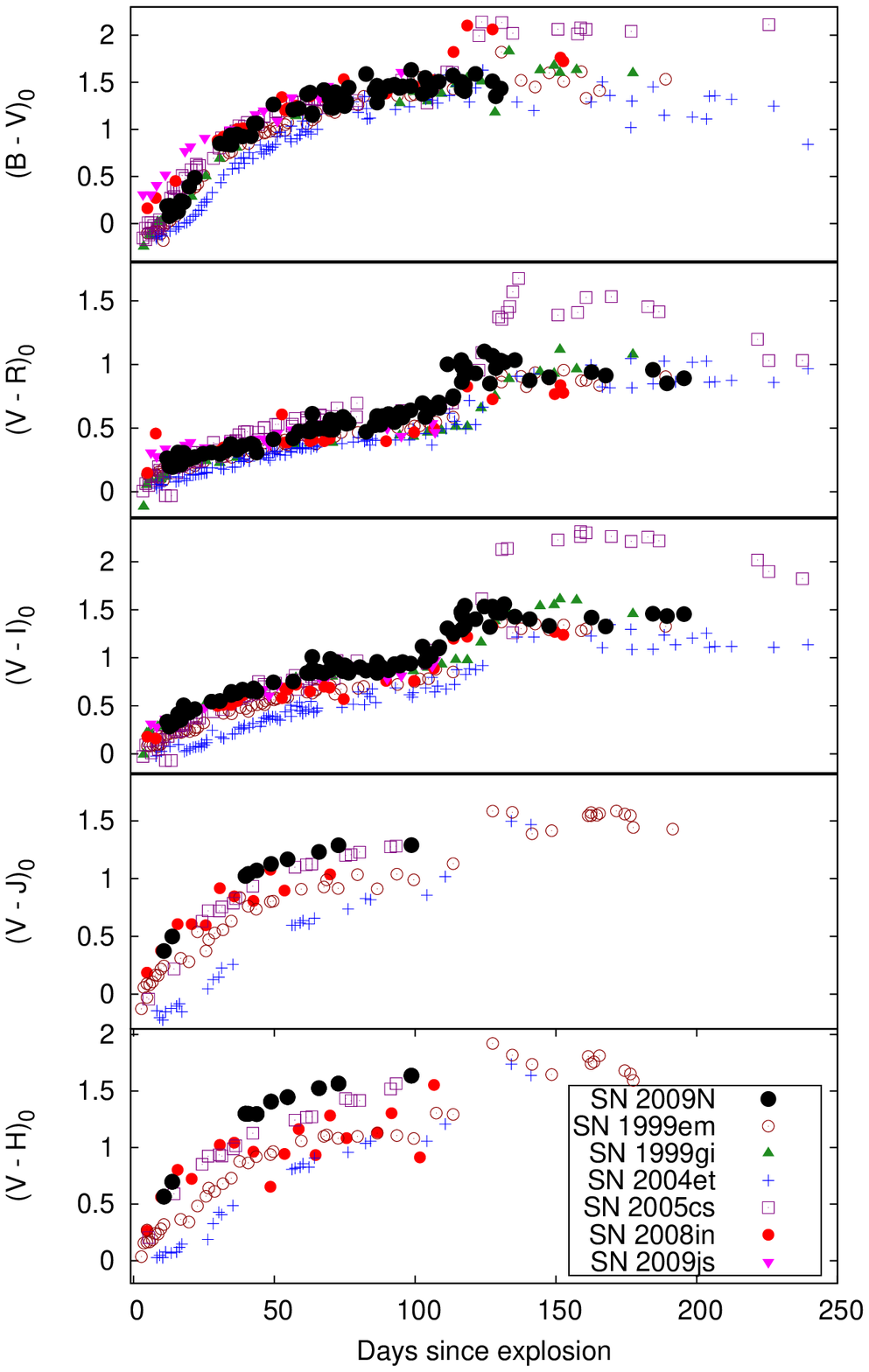}
\caption{Evolution of $(B-V)_0$, $(V-R)_0$, $(V-I)_0$, $(V-J)_0$, and $(V-H)_0$ colours of SN 2009N, together with those of other SNe II-P: SNe 1999em, 1999gi, 2004et, 2005cs, 2008in, and 2009js. The adopted reddening and the sources of photometry of these SNe are in Table \ref{otherSN}.}
\label{colour}
\end{figure}

In Fig. \ref{colour} we show the $(B-V)_0$, $(V-R)_0$, $(V-I)_0$, $(V-J)_0$ and $(V-H)_0$ colour curves of SN 2009N corrected for the reddening derived above, along with data of other type II-P SNe. The extinction and explosion epoch of the compared SNe -- which were chosen to represent a sample with large variety of physical properties -- are summarised in Table \ref{otherSN}. The optical colour evolution of SN 2009N is similar to those of other II-P SNe. The $(B-V)_0$ colour is blue at the beginning and becomes redder quickly, due to the appearance of strong metallic lines in $B$ band. The evolution of the $(V-R)_0$ colour is slower. The $(V-I)_0$ colour of the 6 SNe shows somewhat higher scatter during the plateau phase than in the other two colours. SNe 2005cs and 2008in, along with SN 2009N seem to be redder than the other SNe. This is more pronounced in the near-infrared colours, where SN 2009N is the reddest one in the sample. Since the determination of $E(B-V)$ is uncertain, we cannot rule out that we underestimated its value. However, even with significantly higher extinction the infrared colours of SN~2009N remain one of the reddest ones in the sample. 

\begin{table}
 \centering
 \begin{minipage}{84mm}
  \caption{Explosion epoch, reddening and distance of the SNe used for comparison throughout the paper.}
  \label{otherSN}
  \begin{tabular}{@{}ccccc@{}}
  \hline
  \hline
SN & $t_0$ (JD) & $E(B-V)$ & $D$ (Mpc) & Ref. \\
\hline
1999em & 2451477.0  & 0.10 & 11.7 & 1,2 \\
1999gi & 2451518.3 & 0.21 & 11.1 & 3 \\
2004et & 2453270.5 & 0.41 & 4.8 & 4,5 \\
2005cs & 2453549.0 & 0.05 & 8.4 & 6,7 \\
2008in & 2454825.6 & 0.098 & 13.2 & 8  \\
2009js & 2455115.9 & 0.36 & 21.7 & 9 \\ 
\hline				  
\end{tabular}			  
 \begin{tablenotes}
       \item[a]{References: (1) \citet{hamuy_epm}, (2) \citet{leonard_99em}, (3) \citet{leonard_99gi}, (4) \citet{takats2012},  (5) \citet{maguire_04et},  (6) \citet{pastorello05csII}, (7) \citet{vinko11dh}, (8) \citet{roy08in} (9) \citet{gandhi_09js}}
      
     \end{tablenotes}
\end{minipage}
\end{table}

%###############################   bolometric   ########################################x

\subsection{Bolometric luminosity}\label{sec_bollum}

We calculated the quasi-bolometric luminosity of SN 2009N at those epochs when simultaneous $VRI$ observations were available. If $B$ band data were not taken at these epochs, they were calculated by interpolating the magnitudes from neighboring nights using low-order polynomials. We determined the optical counterpart of the bolometric flux by correcting the observed $BVRI$ magnitudes for reddening, converting them to flux densities at the effective wavelength of the filters \citep{bessell1983} and integrating them using Simpson's rule. The bolometric fluxes were then converted into luminosities using the distance $D=21.6 \pm 1.1$ Mpc (Sec. \ref{sec_avedist}). 

During the first $60$ days of evolution {\it Swift} UV photometry was available (Sec. \ref{sec_swift}). We integrated the UV fluxes at each epoch, then extrapolated the resulting flux curve to the epochs of the optical photometry. Since the UV light curve decreases quickly, we assume, that the UV counterpart of the bolometric flux is marginal ($\sim$ 5\%) at later phases of the plateau (after day $+80$) and negligible during the nebular phase. 
We have NIR photometric observations between days $+10$ and $+98$ after explosion (Sec. \ref{photometry}). We calculated the NIR contribution to the bolometric flux by integrating the fluxes from the red edge of the $I$ band to $H$ band. The flux redwards of $H$ band was approximated with the Rayleigh-Jeans tail.
The resulting infrared luminosities were then interpolated to the epochs of the optical observations.
We calculated the bolometric luminosity curve until day $+98$ by adding the optical, UV and infrared luminosities.

At phases later than $+$98 days we lack both UV and NIR data (except at one epoch at day $+427$). To overcome this problem, we considered the bolometric corrections (BC) of \citet{maguire_04et} and \citet{bersten_hamuy}. Using the data of SNe 1987A, 1999em, 2004et, and 2005cs, \citet{maguire_04et} determined BCs from both $V$ and $R$ bands, as a function of phase. We calculated the bolometric luminosity of SN~2009N from both $R$ and $V$ band photometry, which were then averaged. Comparing these to our previously calculated bolometric luminosity curve, we found that they agree excellently. 
\citet{bersten_hamuy} calculated the bolometric correction as a function of colour during the plateau phase, using the bolometric light curves of 3 SNe (1987A, 1999em, and 2003hn) as well as the models of \citet*{E96} and \citet{D05}. Applying their BCs to the data of SN~2009N, we found that the inferred values somewhat (by $5-10$ per cent) overestimated our bolometric luminosities during the first $98$ days of evolution.
Since the BCs of \citet{maguire_04et} led to luminosities that agreed well with our bolometric luminosities during the first $+98$ days, we used these BCs to calculate the luminosity of SN~2009N at phases later than $+98$ days, when UV and NIR data were not available. We also checked these BCs at the nebular phase by interpolating the $YJH$ magnitudes to the epochs of the $VRI$ measurements at days $+372$, $400$,  and $413$, and integrating the fluxes. The resulting bolometric luminosities agreed well (within $0.1$~dex) with the ones calculated with the BCs of \citet{maguire_04et}.

\begin{figure}
\includegraphics[width=84mm]{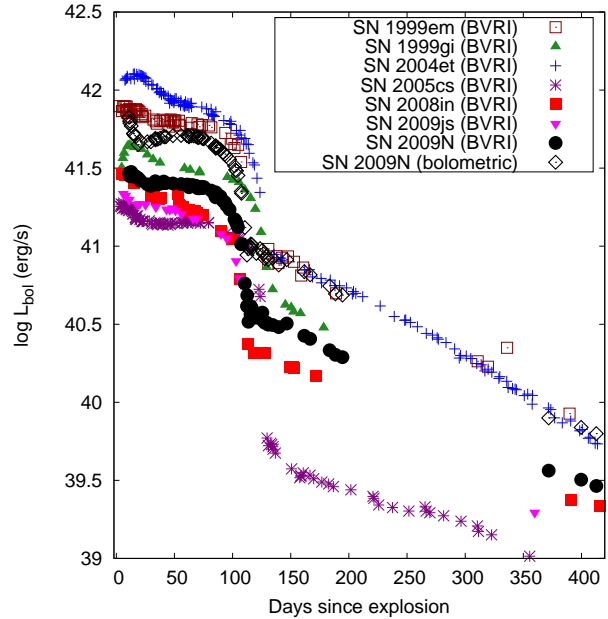}
\caption{Comparison of the evolution of the $BVRI$ quasi-bolometric luminosity of SN 2009N (filled circles) with those of other SNe II-P. The {\it uvoir} bolometric luminosity curve of SN~2009N is also shown (empty diamonds).}
\label{lumosszehas}
\end{figure}

Fig.~\ref{lumosszehas} shows the bolometric luminosity curve of SN~2009N. The observed peak luminosity was $\log L_{\rm bol}=41.82$~erg~s$^{-1}$. Also in Fig.~\ref{lumosszehas}  we compare the  luminosity curve of SN 2009N with those of other SNe II-P. For better comparison we use the $BVRI$ bolometric luminosities for all SNe.
SN 2009N was significantly fainter than the normal SNe II-P 1999em and 2004et during the plateau phase, and about $1.5-1.8$ times brighter than the subluminous SN~2005cs. Its luminosity was comparable, only slightly higher than those of the intermediate luminosity SNe~2008in and 2009js.

%#####################################   spectra ############################################

\section{Spectroscopy}\label{spectroscopy}

\subsection{Optical spectra}

\begin{figure*}
\includegraphics[width=\textwidth]{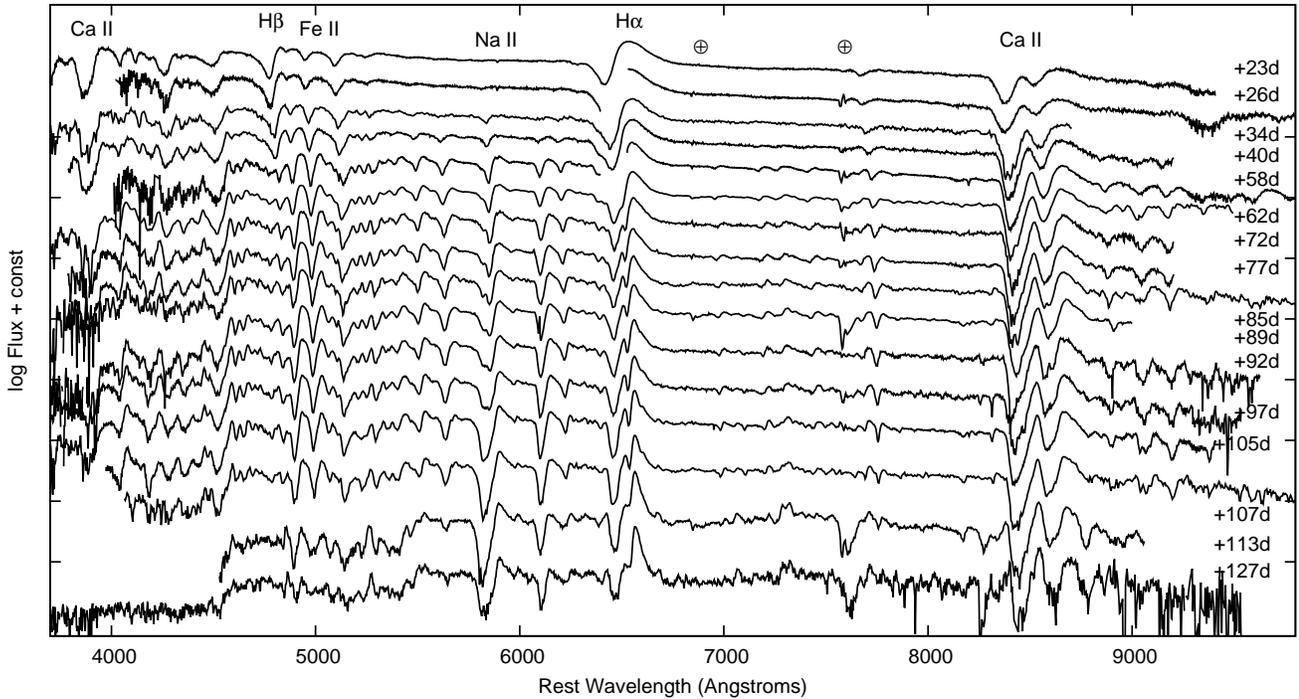}
\caption{Spectral evolution of SN 2009N during the plateau and transition phase. Phases are relative to the estimated explosion date, $t_0= 2454848.1$~JD. The approximate positions of some of the strongest features are marked to guide the eye. The position of telluric features are marked with $\oplus$ symbol.}
\label{sp_seq}
\end{figure*}

Optical spectroscopic observations of SN 2009N covering the phases between days $+$23 and $+$414 after explosion were carried out with multiple telescopes. Table \ref{logofsp} contains the summary of the observations.
The images were reduced and calibrated using standard {\sc iraf} tasks. After bias and flat corrections, the spectra were extracted. The wavelength calibration was carried out using comparison lamp spectra. Standard star spectra taken on the same night were used for the SN flux calibration, which was checked against the photometry at the nearest epoch, and -- when necessary -- the spectra were corrected using a scaling factor.
Optical spectra were also obtained through the CSP, with the Wide Field Reimaging CCD Camera (WFCCD) and Boller \& Chivens Spectrograph (BC)
mounted on the du-Pont, the Low Dispersion Survey Spectrograph (LDSS3) mounted on the Magellan Clay, and the Inamori Magellan Areal Camera and Spectrograph (IMACS) mounted
on the Magellan Baade telescopes at Las Campanas Observatory (Table \ref{logofsp}). Again, overall
survey techniques can be found in \citet{hamuy_CSP}. Spectra were processed
through standard techniques of reduction, extraction, and calibrations, 
a detailed
description of which can be found in \citet{folatelli_CSP}, as applied to
the SN~Ia spectroscopic sample.
We also obtained a medium-resolution spectrum of SN~2009N with the MagE spectrograph \citep{Marshall08} on the Magellan Clay 6.5~m telescope.  For the MagE spectrum, the sky was subtracted from the images using the method described by \citet{Kelson03}.  We employed our own IDL routines for flux calibration and telluric line removal using the well-exposed continua of spectrophotometric standard stars \citep{Wade88, Foley03, Foley09:08ha}.
All plateau and transition phase optical spectra are shown in Fig. \ref{sp_seq}.

 \begin{table*}
 \centering
 \begin{minipage}{\textwidth}
  \caption{Summary of the optical spectroscopic observations.}
  \label{logofsp}
  \begin{tabular}{@{}cccccc@{}}
  \hline
  \hline
Date & JD & Phase$^a$ & Instrument set-up & Wavelength range &  Resolution$^b$ \\
     & 2400000+ & days & & \AA & \AA \\
  \hline
08/02/2009 & 54870.8 & 22.7  & Magellan Clay + LDSS-3 + VPH-All & 3700-10000 & 7 \\
09/02/2009 & 54871.8 & 23.7  & Magellan Clay + LDSS-3 + VPH-All & 3700-10000  & 7 \\
11/02/2009 & 54873.8 & 25.7  & Magellan Baade + IMACS + Gri-200-15.0 & 3900-10000 & 5 \\
18/02/2009 & 54881.6 & 33.5  & 2.2m Calar Alto + CAFOS+blue-200 & 3400-8900 & 12 \\
19/02/2009 & 54882.6 & 34.5  & 2.2m Calar Alto + CAFOS+blue-200 & 3400-8900 & 12 \\
24/02/2009 & 54886.8 & 38.7  & du Pont + WFCCD + Blue grism  & 3800-9100 &  8 \\
25/02/2009 & 54887.9 & 39.8  & du Pont + WFCCD + Blue grism  & 3800-9100 &  8 \\
26/02/2009 & 54888.8 & 40.7  & du Pont + WFCCD + Blue grism  & 3800-9100 &  8  \\
15/03/2009 & 54905.8 & 57.7  & Magellan Baade + IMACS + Gri-200-15.0 & 3900-10000 & 5  \\
19/03/2009 & 54910.5 & 62.4  & NOT + ALFOSC+grism-4 & 3200-9100 & 19 \\
19/03/2009 & 54910.5 & 62.4  & 2.2m Calar Alto + CAFOS+green-200 & 3790-10000 & 12 \\
29/03/2009 & 54919.8 & 71.7  & du Pont + WFCCD + Blue grism  & 3800-9100 &  8  \\
03/04/2009 & 54924.7 & 76.6  & du Pont + WFCCD + Blue grism  & 3800-9100 &  8  \\
11/04/2009 & 54933.5 & 85.4  & 2.2m Calar Alto + CAFOS+green-200 & 3790-10000 & 12  \\
12/04/2009 & 54933.6 & 85.5  & NOT + ALFOSC+grism-4 & 3200-9100 & 19 \\
12/04/2009 & 54934.5 & 86.4  & 2.2m Calar Alto + CAFOS+green-200 & 3790-10000 & 12 \\
16/04/2009 & 54937.5 & 89.4  & NOT + ALFOSC+grism-4 & 3200-9100 & 16 \\
18/04/2009 & 54939.7 & 91.6  & du Pont + BC + 300l/mm grating & 3500-9600 & 8  \\
23/04/2009 & 54944.8 & 96.7  & du Pont + BC + 300l/mm grating & 3500-9600 & 8  \\
01/05/2009 & 54952.7 & 104.5 & Magellan Clay + LDSS-3 + VPH-All & 3700-10000  & 7   \\
03/05/2009 & 54955.4 & 107.3 & 2.2m Calar Alto + CAFOS+green-200 & 3790-10000 & 11 \\
09/05/2009 & 54961.4 & 113.3 & NOT + ALFOSC+grism-4 & 3200-9100 & 15 \\
23/05/2009 & 54974.6 & 126.5 & du Pont + BC + 300l/mm grating & 3500-9600 & 8 \\ 
24/06/2009 & 55006.5 & 158.4 & Magellan Clay + MagE &  3210-10350 & 1-2 \\
25/01/2010 & 55221.8 & 373.7 & NTT + EFOSC2+gr\#16 & 6000-10100 & 14 \\
06/03/2010 & 55261.8 & 413.7 & VLT + FORS2+300V & 3500-9600 & 10 \\
\hline				  
\end{tabular}			  
 \begin{tablenotes}
       \item[a]{$^a$ relative to the estimated date of explosion, $t_0=2454848.1$~JD}
       \item[b]{$^b$ as measured from the full-width at half maximum (FWHM) of the night-sky lines}
     \end{tablenotes}
\end{minipage}
\end{table*}

\begin{figure*}
\includegraphics[width=\textwidth]{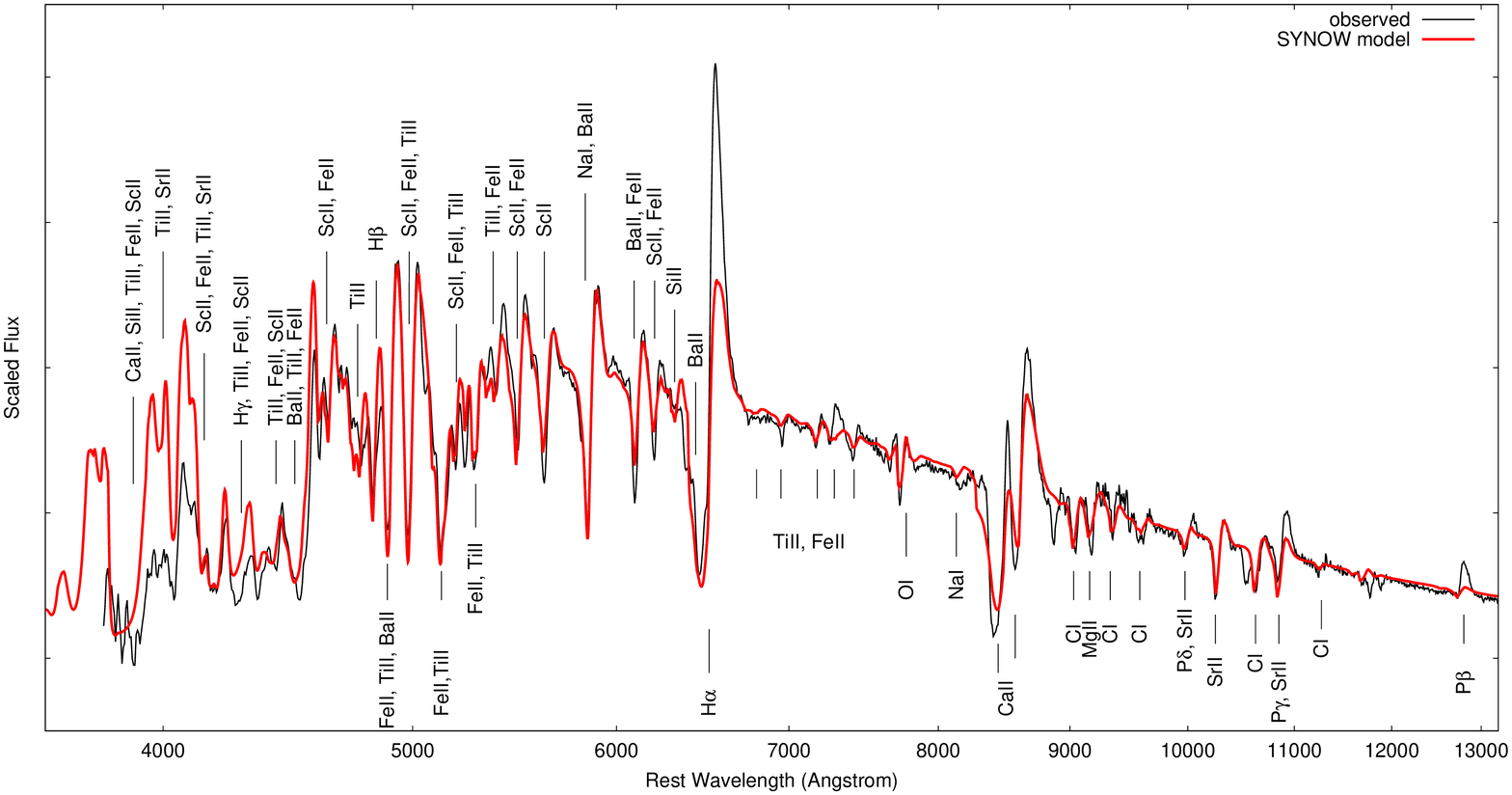}
\caption{{\sc synow} model of the reddening and redshift ($z=0.0035$) corrected, combined optical and near-infrared spectra, taken on days $+$62 and $+$59 with the 2.2~m Calar Alto Telescope and with NTT, respectively.}
\label{synow}
\end{figure*}

The classification spectrum was taken $\sim$8 days after explosion by \citet{09N_classif}. It has low signal-to-noise ratio, but the Balmer series of H\,{\sc i} are clearly visible\footnote{\url{http://www.cfa.harvard.edu/supernova/spectra/sn2009N.gif}}. Using the SNID \citep{SNID}, the spectrum was found to be similar to that of SN~2005cs two days after its maximum.
Our earliest spectra of SN 2009N were taken 23, 24, and 26 days after the explosion. At that time the metallic lines already appeared in the spectrum. Next to the strong, wide Balmer series of H\,{\sc i}, features of Fe\,{\sc ii} and Ca\,{\sc ii} were present. 
By day $+$34 features of Na\,{\sc i}, O\,{\sc i}, Si\,{\sc ii}, Ti\,{\sc ii}, Sc\,{\sc ii} also appeared, and remained visible throughout the plateau phase. The Na\,{\sc i} D feature was very weak at that point, but became stronger at later phases. 

\begin{figure}
\includegraphics[width=84mm]{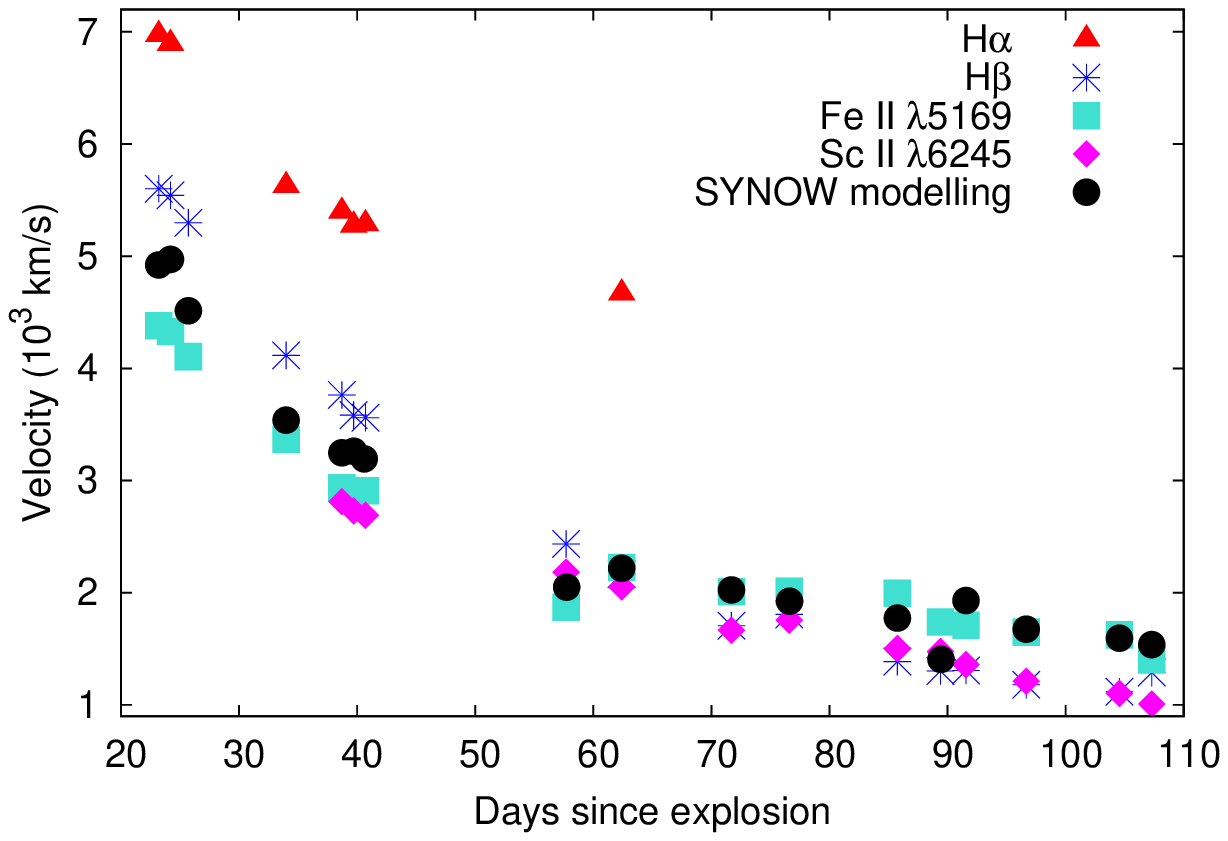}
\caption{Expansion velocities measured from the absorption minima of selected lines, together with the photosperic velocities determined via {\sc synow} modelling.}
\label{line_vel}
\end{figure}

\begin{figure}
\includegraphics[width=84mm]{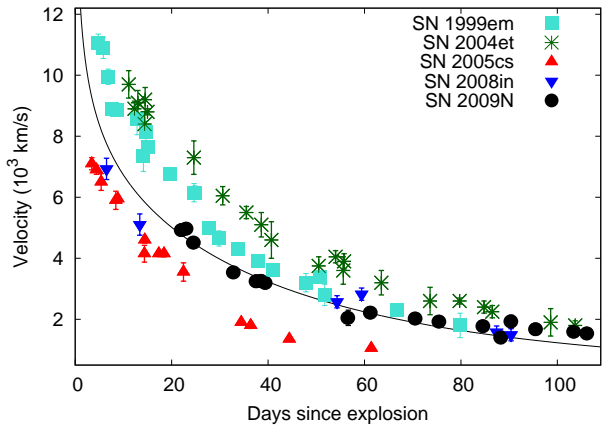}
\caption{Comparison of the velocities of SN~2009N determined via {\sc synow} modelling with those of other SNe II-P \citep{takats2012}. In order to determine the distance via EPM, the velocities of SN~2009N were extrapolated to the epochs of the photometry using the relation of \citet{takats2012} (solid line; see Sec. \ref{epm}).}
\label{vel_osszehas}
\end{figure}

We created models with {\sc synow} \citep{fisher,hatano} for the plateau phase spectra for line identification. The method described by \citet{takats2012} was applied to find the set of parameters of the model that fits best the observed spectrum and to measure the photospheric velocity. An example model with line identification is shown in Fig. \ref{synow}.
We also measured the expansion velocities of selected lines by fitting a Gaussian to their absorption components and determining the blueshift of the minima.
The line velocities of H$\alpha$, H$\beta$, Fe\,{\sc ii}~$\lambda$5169, and Sc\,{\sc ii}~$\lambda$6245 are shown in Fig. \ref{line_vel} together with the photospheric velocities determined via {\sc synow} modelling.  The photospheric velocities of SN~2009N are between those of normal and subluminous SNe (Fig. \ref{vel_osszehas}).

\begin{figure}
\includegraphics[width=84mm]{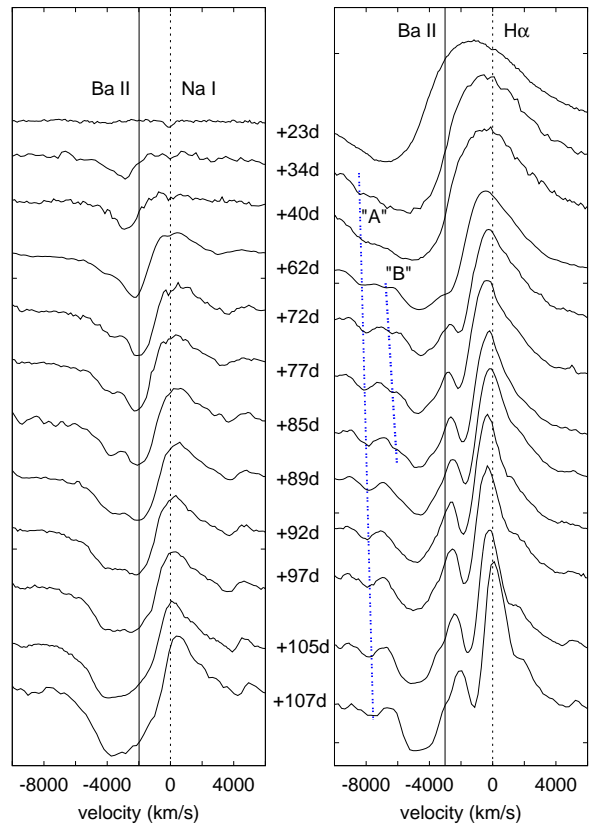}
\caption{Evolution of the Na\,{\sc i} D and Ba\,{\sc ii} $\lambda$5854 (left) and H$\alpha$ and Ba\,{\sc ii} $\lambda$6497 (right) lines. The black vertical lines show the position of the rest wavelength of these lines. The blue dotted lines in the right panel, marked ``A'' and ``B'' mark the positions of the shallow absorption features that may be high-velocity components of H$\alpha$ (see also Sec.~\ref{sec_nir}).}
\label{line_evol}
\end{figure}

An interesting detail is the appearance and strengthening of the Ba\,{\sc ii} $\lambda\lambda$5854 and 6497 lines. In Fig. \ref{line_evol} we enlarged the wavelength ranges around Na\,{\sc i} D and H$\alpha$ to show the evolution of these Ba\,{\sc ii} features. The Ba\,{\sc ii} lines are not visible in the first two spectra, and our {\sc synow} models do not show any need of their presence either. On day $+$40, both Ba\,{\sc ii} $\lambda\lambda$5854 and 6497 lines are very weak, but detectable, together with Ba\,{\sc ii} $\lambda$6142 (Fig.~\ref{sp_seq}). By day $+$62 the Ba\,{\sc ii} $\lambda$6497 line is clearly visible next to H$\alpha$, while Ba\,{\sc ii} $\lambda$5854 is weak, but noticeable next to Na\,{\sc i}.
Both features become stronger and stronger throughout the plateau phase. Ba\,{\sc ii} $\lambda$5854 forms a blend with Na\,{\sc i}, but its contribution to the line shape can be confirmed with {\sc synow}. These strong Ba\,{\sc ii} features seem to be typical of subluminous II-P SNe: they were detected in the spectra of e.g. SNe 2005cs \citep{pastorello05csII}, 2008in \citep{roy08in}, and 2009md \citep{fraser09md}.

\begin{figure*}
\includegraphics[width=\textwidth]{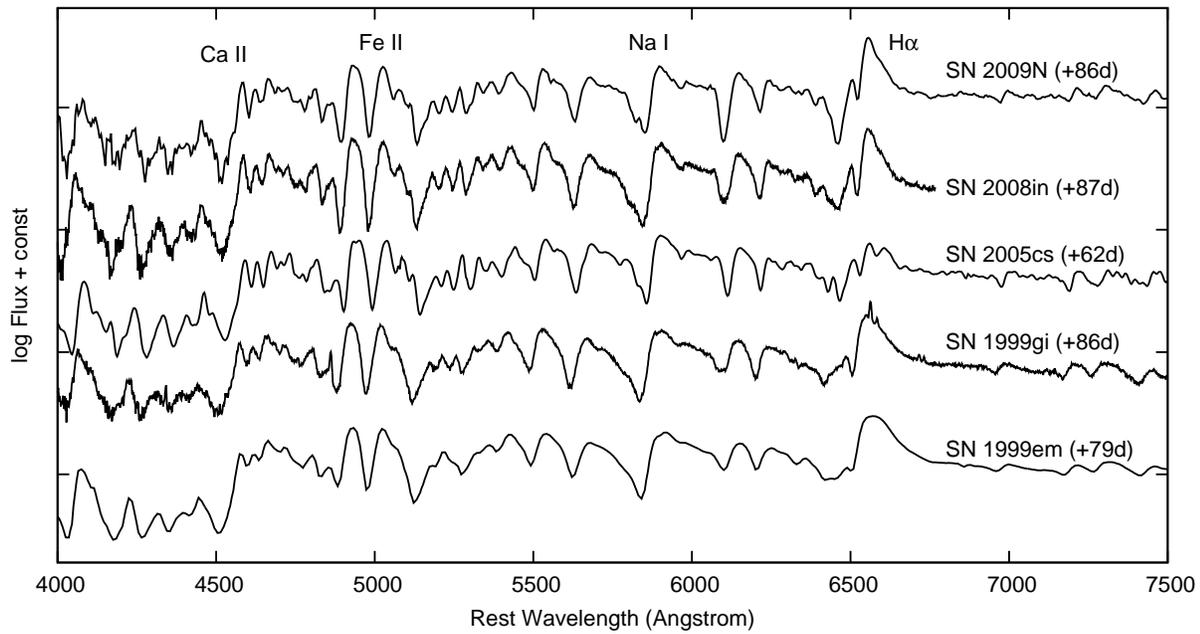}
\caption{Comparison of the spectra of SNe 2009N and 2008in \citep{roy08in} with the subluminous SN 2005cs \citep{pastorello05csII} and the normal type II-P SNe  1999em \citep{leonard_99em} and 1999gi \citep{leonard_99gi}. Some of the strongest features are labeled to help the eye.}
\label{sp_osszehas}
\end{figure*}

By modelling the spectra of SN 1997D, \citet{turatto_ba} showed that the appearance of the strong Ba\,{\sc ii} lines is not due to the overabundance of Ba, an s-process element, but it is likely a temperature effect. 
For several elements and supernova atmosphere compositions \citet{hatano} examined how the local thermodynamic equilibrium (LTE) optical depth of one of the strongest lines of each element changes with the temperature. Their Fig. 2 shows that Ba\,{\sc ii} lines appear bellow $\sim 6000$~K, and the optical depth of the Ba\,{\sc ii} reference line increases very quickly as the temperature decreases. 
This agrees well with the evolution of Ba\,{\sc ii} features observed in SN 2009N. The temperature decreases to $6000$~K around day $+$29 (Table \ref{epm_data}, see also Sec. \ref{epm}), and Ba\,{\sc ii} cannot be detected in the spectra taken on days $+$34 and $+$35. By the epoch of the next spectra (day $+$39, $+$40, $+$41), the temperature drops to $\sim 5700$~K, and the Ba\,{\sc ii} lines start to appear, and later, with the temperature decreasing further, the features become more and more pronounced.
The fact that Ba\,{\sc ii} $\lambda 6497$ seems to appear on the day $+$79 spectrum of SN 1999em (see Fig. \ref{sp_osszehas}), when its temperature dropped under $6000$~K \citep[see Table 7 of][]{leonard_99em}, is also in agreement with the behavior of Ba\,{\sc ii} in SN 2009N.

In addition to their lower temperatures, subluminous SNe have lower expansion velocities, narrower spectral features, which makes Ba\,{\sc ii} lines visible at earlier phases and easier to identify.

In Fig. \ref{sp_osszehas} we compare the spectrum of SN 2009N taken on day $+$86 to the spectra of SNe 2008in, 1999gi, 2005cs, and 1999em at similar phases. The spectrum of SN 2008in -- which has luminosities similar to SN 2009N (Fig. \ref{lumosszehas}) -- well matches that of SN~2009N; they also show similarities to that of the subluminous SN 2005cs. The spectra of the normal type II-P SNe 1999gi and 1999em are different: the features are wider and implying higher ejecta velocities.

\begin{figure*}
\includegraphics[width=\textwidth]{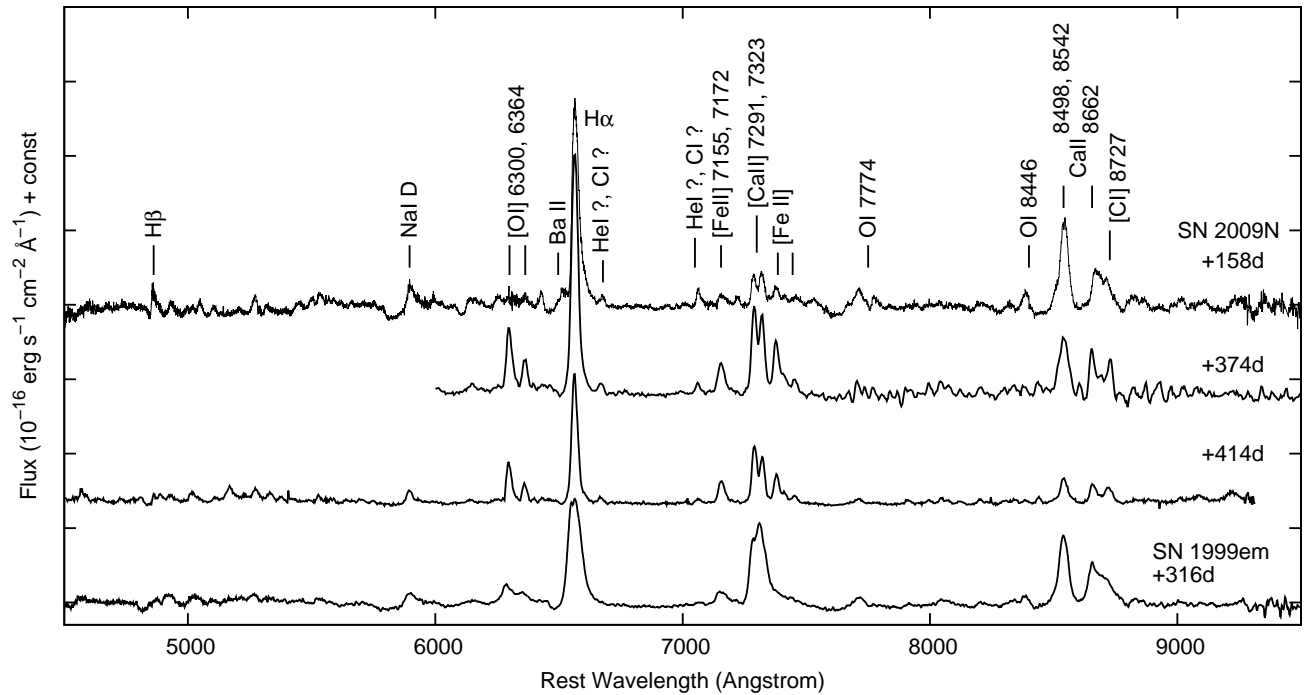}
\caption{Nebular phase spectra of SN 2009N obtained on days $+$159, $+$374, and $+$414. For comparison we plotted the spectrum of SN 1999em taken on +316d. \citep{leonard_99em}.}
\label{sp_seq_nebu}
\end{figure*}

The nebular phase spectra of SN 2009N taken on days $+$158, $+$374, and $+$414, are shown in Fig. \ref{sp_seq_nebu}. The most dominant features are H$\alpha$ and the Ca\,{\sc ii} IR triplet. On day $+$158 next to H$\alpha$ and Na\,{\sc i} D, the lines of Ba\,{\sc ii} are still strong and have P Cygni line profiles. The spectra taken on days $+$374 and $+$414 show only minor differences. They are similar to nebular phase spectra of other type II-P SNe, such as SN~1999em, but with narrower features. Next to the strong H$\alpha$ emission line and the Ca\,{\sc ii} IR triplet, features of O\,{\sc i} and several forbidden lines can be identified, such as the [O\,{\sc i}] 6300, 6364 \AA~doublet, [Fe\,{\sc ii}] 7155, 7273, 7439 \AA, [Ca\,{\sc ii}] 7291,7323 \AA~doublet, [C\,{\sc i}] 8727\AA. 
 [Fe\,{\sc ii}] 7273 \AA~appears to be somewhat stronger than in the spectra of subluminous SNe \citep[e.g.][]{benetti_1997D, pastorello05csII}.

Our two latest spectra were included in the sample of \citet{maguire_nebusp}, who studied the nebular spectra of type II-P SNe. Measuring the velocities from the width of the emission lines and comparing them to those of other SNe, they found that the nebular phase velocities of SN 2009N are quite low ($400 - 700$ km~s$^{-1}$), and similar to those of the subluminous SN~2005cs and SN~2008bk.
\citet{maguire_nebusp} also noted that the  [OI] 6300 \AA~line profile is somewhat blueshifted at both epochs, which can be a sign of dust formation in the ejecta (see their Fig. 3).

\subsection{Near infrared spectra}\label{sec_nir}

Near-infrared spectra were taken at five epochs during the plateau phase with three different instruments, NTT/SOFI, VLT/ISAAC, and SOAR/OSIRIS covering the phases between days $+$17 and $+$63 after explosion. The summary of the observations can be found in Table \ref{logofnirsp}. The reduction of the spectra was carried out using {\sc iraf} packages. At each epoch several pairs of spectra were taken at different positions along the slit. The pairs of images were subtracted from each other to remove the sky background. These subtracted images were then added together. The SN spectrum was extracted from the co-added image. The wavelength calibration was carried out using arc lamp spectra taken at the same night as the SN spectrum. We removed the strong telluric features using the spectrum of a G-type telluric standard star observed close in time and at similar airmass as the SN. 
The spectrum of the telluric standard was also used for the flux calibration of the SN spectra that were later checked against the NIR photometry from the nearest epoch and corrected when necessary.

\begin{figure}
\includegraphics[width=84mm]{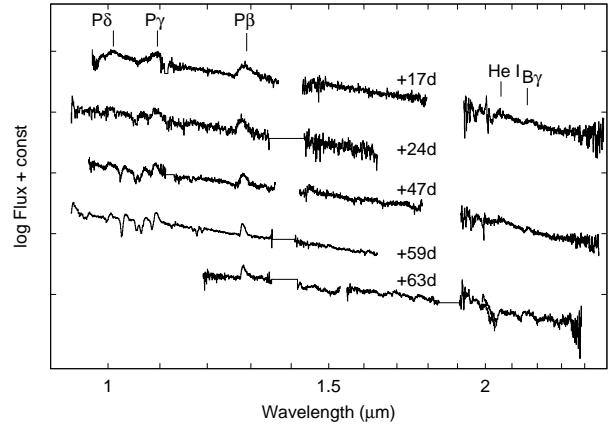}
\caption{Near infrared spectra of SN 2009N. The phases are relative to the estimated date of explosion, $t_0 = 2454848.1$~JD.}
\label{nir_seq1}
\end{figure}

\begin{figure}
\includegraphics[width=84mm]{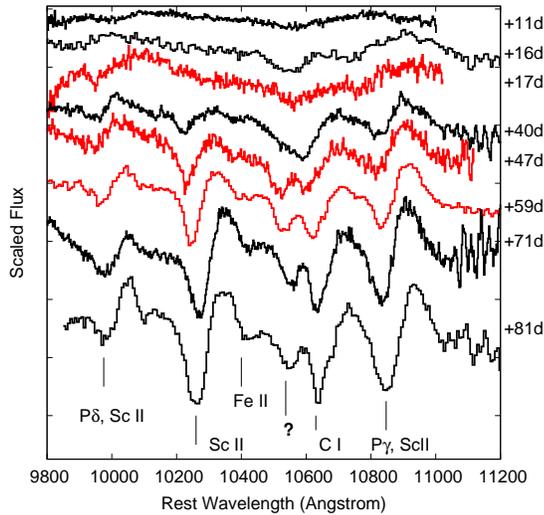}
\caption{The emergence and evolution of the feature at about 1.055$\mu{\rm m}$ (marked with ``?'') in the NIR spectra of SNe~2009N (red) and 2008in (black).}
\label{nir_seq}
\end{figure}

The sequence of the NIR spectra of SN 2009N is presented in Fig.~\ref{nir_seq1}. In the first spectrum, obtained on day $+$17  after the explosion, the Paschen series of H\,{\sc i} as well as the He\,{\sc i} feature at 2.058 $\mathrm{\mu m}$ are visible. By day $+$24, the strong C\,{\sc i} $\lambda$10691 line is recognizable next to P$\gamma$. At later phases lines of Fe\,{\sc ii}, Sr\,{\sc ii} and Ca\,{\sc ii} can be identified (Fig. \ref{nir_seq}).

\begin{figure}
\includegraphics[width=84mm]{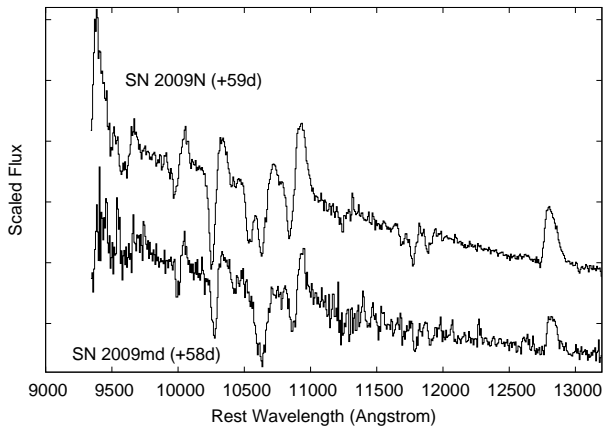}
\caption{Comparison of the NIR spectrum of SN~2009N at $+$59 days after explosion with that of SN~2009md around the same phase \citep{fraser09md}.}
\label{comp_09md}
\end{figure}

Next to C\,{\sc i} $\lambda$10691, at about $1.055\,{\rm\mu m}$, an absorption feature emerged by day $+47$, which is not present in the spectra of normal type II-P SNe. The reason for this could be that such objects have higher velocities and broader spectral lines, therefore since this wavelength range is dominated by the C\,{\sc i}  line, the two features -- C\,{\sc i} and the the one at $\sim1.055\,{\rm\mu m}$ -- may be blended. The sample of subluminous SNe that have NIR spectra during the second half of the plateau phase is quite small, and none of them have this feature present. In Fig. \ref{comp_09md} we compare the NIR spectrum of SN~2009N taken at $+$59 days to that of SN~2009md \citep{fraser09md} at a similar phase. SN~2009md was a subluminous SN, with narrow, low-velocity spectral lines, and its NIR spectrum is similar to that of SN~2009N, but the line at $1.055\,{\rm\mu m}$ is not present. 

Looking at our previously unpublished NIR spectra of SN 2008in (Appendix \ref{app_a}), however, we found, that they showed the same feature. In Fig. \ref{nir_seq} the NIR spectra of the two SNe are plotted together, showing the evolution of this feature over time.

In order to try to identify the possible transition that produced this feature, we combined the optical spectrum taken on day $+62$ with the NIR spectrum on day $+59$, and used {\sc synow} modelling. We have found the following two possibilities.
 
First, we managed to model this feature at $\sim1.055\,{\rm\mu m}$ with a high-velocity (HV) component of He\,{\sc i}~$\lambda$10830, having the velocity of $\sim$8000 km~s$^{-1}$. 
By modelling the interaction between the ejecta and the circumstellar material (CSM) originated from average RSG wind, \citet*{chugai_intera} have shown that due to excitation of the unshocked ejecta HV features of He\,{\sc i}~$\lambda$10830 and H$\alpha$ can emerge \citep[see also][]{inserra2012, inserra_sne}. We examined the optical spectra of SN 2009N (Sec. \ref{spectroscopy}) to see if a HV component of H$\alpha$ was also present. As Fig. \ref{line_evol} shows, a shallow absorption feature is visible at $\sim$8000 km~s$^{-1}$ (marked as ``A''). It appears around day $+$39 and is present during the entire plateau phase. 
A smaller dip is also visible next to H$\alpha$ between days $+$62 and $+$77 (marked ``B'' in Fig.~\ref{line_evol}). If it is a HV component of H$\alpha$, its velocity is significantly lower and decreases more quickly. We also examined the region of H$\beta$, and found several weak absorption features, some of them at similar, but somewhat lower velocities as those near H$\alpha$. However, since there are many metallic lines in this region, we cannot claim for certain that any of them is a HV component of H$\beta$. There is no visible HV component of He\,{\sc i} $\lambda$5876 either. 
The optical spectra of SN 2008in also show similar lines near H$\alpha$ as in the case of SN~2009N. In Fig.~8 of \cite{roy08in} an absorption feature (marked ``C'') is visible. The authors assumed it to be originated from Fe\,{\sc ii} multiplets, but they also noted that the presence of HV H$\alpha$ component could not be ruled out.

\begin{figure}
\includegraphics[width=84mm]{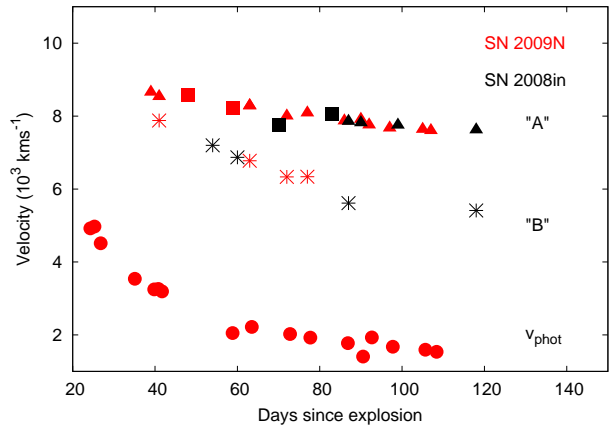}
\caption{The velocities of the (assumed) high-velocity features in the spectra of SNe~2009N (red) and 2008in (black): the HV H$\alpha$ feature ``A'' (triangles, see Fig. \ref{line_evol}), the HV He\,{\sc i} (squares) and the HV H$\alpha$ feature ``B'' (asterisks). The photospheric velocities of SN~2009N, determined via {\sc synow} modelling are also shown for comparison (filled circles).} 
\label{HV_abra}
\end{figure}

We measured the velocities of the assumed HV features next to H$\alpha$ and He\,{\sc i} $\lambda$10830 in the case of both SNe 2009N and 2008in. The values are shown in Fig. \ref{HV_abra}. The velocities of HV H$\alpha$ (marked "A" in Fig. \ref{line_evol}) and HV He\,{\sc i}~$\lambda$10830 are consistent and decrease very slowly, which supports the CSM interaction scenario. However, it seems somewhat strange that these velocities agree so well for SNe 2009N and 2008in, which suggests very similar pre-supernova evolution. For both SNe there is also a weaker feature present next to H$\alpha$ with lower velocities that decrease more quickly.

\begin{figure}
\includegraphics[width=84mm]{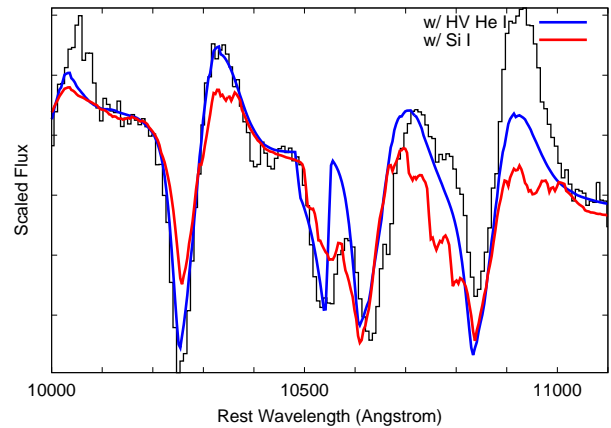}
\caption{The wavelength range around the feature at $\sim1.055{\rm \mu m}$ on day $+59$, together with the best-fitting {\sc synow} models including either Si\,{\sc i} (red) or HV He\,{\sc i} (blue).}
\label{models}
\end{figure}

Alternatively the feature at $1.055\,{\rm\mu m}$ can also be modelled with Si\,{\sc i}. However, due to the fact that Si\,{\sc i} has several lines in the range between $1.07-1.09\,{\rm \mu m}$ that are not present in the observed spectra, our {\sc synow} models were unable to fit the depth of the absorption feature (Fig. \ref{models}).  As in the case of the Ba\,{\sc ii} features, the presence of Si\,{\sc i} during the second half of the plateau phase is probably a temperature effect.

We plotted the observed spectrum together with the best-fitting models in both cases, i.e. including Si\,{\sc i} or HV He\,{\sc i} in Fig. \ref{models}, showing that the model with HV He\,{\sc i} fits the observed spectrum better.

 \begin{table*}
 \centering
 \begin{minipage}{\textwidth}
  \caption{Summary of the near infrared spectroscopic observations.}
  \label{logofnirsp}
  \begin{tabular}{@{}ccccccc@{}}
  \hline
  \hline
Date & JD & Phase$^a$ & Instrument &  Wavelength range &  Resolution \\
     & 2400000+ & days & & ${\rm\mu m}$ &   \\
  \hline
03/02/2009 & 54865.5 & 17.4 & VLT+ ISAAC (SWS1-LR) & 0.98-2.5 & 500\\
09/02/2009 & 54872.5 & 24.4 & NTT + SOFI (GB,GR) & 0.95-1.64, 1.53-2.52 & 1000 \\
04/03/2009 & 54895.5 & 47.4 & VLT+ISAAC (SWS1-LR) & 0.98-2.5 & 500 \\
16/03/2009 & 54907.5 & 59.4 & NTT+SOFI (GB,GR) & 0.95-1.64, 1.53-2.52 & 1000 \\
20/03/2009 & 54911.4 & 63.3 & SOAR+OSIRIS (Low-Res) & 1.0-2.58 & 1200 \\
\hline				  
\end{tabular}			  
 \begin{tablenotes}
       \item[a]{$^a$ relative to $t_0=2454848.1$~JD.}
     \end{tablenotes}
\end{minipage}
\end{table*}

%######################################### distance   ###############################################x

\section{Distance}\label{distance}

The estimation of the physical properties of SNe depends on the knowledge of their distance. In the case of SNe II-P different distance measurement techniques have been developed. One of them is the expanding photosphere method  \citep[EPM,][]{epm_ref}, a variant of the Baade-Wesselink method, which requires photospheric and spectroscopic monitoring throughout the first half of the plateau phase, but does not need external calibration. Another one is the standardized candle method \citep[SCM,][]{hamuy_scm}, which is based on the correlation between the SN brightness and the expansion velocity in the middle of the plateau phase. It needs less input data, but requires calibration via SNe with well-known distances.
In this section we determine the distance of SN 2009N applying both mentioned methods, and then we compare and discuss the results.

\subsection{Expanding Photosphere Method}\label{epm}

Using the assumption that at early phases the SN has optically thick, homologously expanding ejecta, which radiates as a diluted blackbody, the expanding photosphere method (EPM) derives its distance by relating the apparent angular size of the photosphere to its physical radius \citep[see for details e.g.][]{hamuy_thesis, leonard_99em, D05}. By measuring the photospheric velocity and determining the angular radius $\theta$ from photometric observations on multiple epochs, the parameters $t_0$ and $D$ can be derived by fitting the linear equation of $t=D \cdot (\theta/v_{\rm phot}) + t_0$.

We used {\sc synow} as described in \citet{takats2012} to model the observed spectra (see also Sec.~\ref{spectroscopy}) and to determine the photospheric velocities. Since we have good photometric coverage, but not so many spectra taken during the first $50$ days of evolution, we extrapolated the velocities to the epochs of the photometric measurements using the formula given in Eq. 3 of \citet{takats2012}. Both $t_0$ and the photospheric velocity on day $+50$ ($v_{50{\mathrm d}}$, which also depends on $t_0$) were fitting parameters, and the values of $t_0=2454849.4 \pm 5.7$~JD and $v_{50{\mathrm d}, {\rm phot}}= 2588 \pm 675$ km~s$^{-1}$ were obtained. The values of the extrapolated velocities are in Table \ref{epm_data} (see also Fig.~\ref{vel_osszehas}).

We used two slightly different approaches to calculate the angular radius ($\theta$). 
First, the  method of \citet{hamuy_epm} was applied by minimizing the quantity of:
\begin{equation}
 \chi^2=\sum_{\lambda}{{[m_{\lambda}+5 \log (\theta \zeta(T)) -b_{\lambda}(T)] \over \sigma_m^2}},
\end{equation}
where $m_\lambda$ is the dereddened apparent magnitude in the filter with the central wavelength $\lambda$, $\sigma_m$ is the photometric error of $m_\lambda$, $b_\lambda(T)$ is the syntetic magnitudes of the blackbody flux at temperature $T$, $\zeta(T)$ is the flux dilution factor. In this way both $\theta$ and $T$ are determined simultaneously. The $BVI$ filter combination was taken into account. We refer to this version as ``multicolour''.

We also applied EPM in a ``bolometric'' way \citep{vinko_02ap}. In this case the angular radius was calculated as
\begin{equation}
\theta= \sqrt{{F_{bol} \over \zeta^2(T) \sigma T^4}}
\end{equation}
We used the bolometric fluxes determined in Sec. \ref{sec_bollum} ($F_{bol}$), while the temperatures were obtained by fitting blackbody curves to the fluxes measured in the $BVRIJH$ bands.

To calculate the distance, we adopted the $\zeta(T)$ dilution factors of \citet{D05}.
Table \ref{epm_data} contains the derived $\theta$, $T$ and $\zeta$ values for both methods along with the extrapolated $v_{\rm phot}$ velocities.

Using the ``multicolour'' method the inferred distance is $D_{\rm m}=22.96 \pm 2.85$~Mpc, while for the epoch of explosion we obtained $t_{0,{\rm m}}=2454847.0 \pm 2.2$~JD. With the ``bolometric'' version the results are $D_{\rm b}=23.64 \pm 2.65$~Mpc and $t_{0,{\rm b}}=2454848.9 \pm 1.5$~JD. We adopted the weighted mean of the results of the two versions of EPM, $D_{EPM}=23.31 \pm 1.94$~Mpc ($\mu =31.83 \pm 0.18$~mag) and $t_0=2454848.1 \pm 1.2$~JD as the EPM distance to SN 2009N and the epoch of its explosion (which date we use throughout the paper), respectively.

The distances determined via EPM depend heavily on the $\zeta(T)$ dilution factors that are calculated from atmosphere models. Currently there are two sets of models that can be used for this purpose, published by \citet{E96} and \citet{D05}. The $\zeta(T)$ curves derived from these two model sets, however, differ significantly, those of \citet{E96} being systematically lower, therefore leading to lower distances \citep[as was discussed in][]{D05}. In the case of SN 2009N the difference in the distance is $\sim14$ per cent.

 \begin{table*}
 \centering
 \begin{minipage}{170mm}
  \caption{Quantities derived in EPM: angular size ($\theta$), temperature (T) and dilution factors ($\zeta$) from the models of \citet{D05} along with the extrapolated velocities.}
  \label{epm_data}
  \begin{tabular}{@{}ccccccccccc@{}}
  \hline
  \hline
 & \multicolumn{3}{c}{``bolometric''} & & \multicolumn{3}{c}{``multicolour''} &  extrapolated velocities \\
JD &  $\theta$ & T &  $\zeta$ & &  $\theta$ & T  & $\zeta$ &  $v_{\rm phot}$  \\
(2400000+) &  ($10^8$ ${\mathrm{km}\over \mathrm{Mpc}}$) & (K) &  & &  ($10^8$ ${\mathrm{km}\over \mathrm{Mpc}}$) & (K) &  & (km~s$^{-1}$)  \\
\hline
54858.9 & 3.20 & 9006  & 0.540 & & 3.47 & 8998 & 0.588 & 6843 \\
54859.7 & 2.48 & 10042 & 0.431 & & 3.15 & 9935 & 0.548 & 6659 \\
54859.8 & 2.93 & 9252  & 0.511 & & 3.38 & 9050 & 0.589 & 6638 \\
54860.9 & 2.50 & 9996  & 0.451 & & 3.44 & 9152 & 0.622 & 6403 \\
54862.8 & 2.84 & 8988  & 0.546 & & 3.51 & 8737 & 0.676 & 6017 \\
54862.8 & 2.78 & 9137  & 0.535 & & 3.71 & 8192 & 0.714 & 6021 \\
54863.8 & 2.96 & 8691  & 0.585 & & 3.63 & 8505 & 0.719 & 5843 \\
54864.8 & 3.17 & 8212  & 0.646 & & 3.96 & 7768 & 0.807 & 5675 \\
54866.8 & 3.50 & 7521  & 0.754 & & 3.99 & 7569 & 0.861 & 5370 \\
54868.8 & 3.68 & 7017  & 0.836 & & 4.11 & 7188 & 0.935 & 5095 \\
54874.7 & 5.03 & 4401  & 1.318 & & 4.37 & 6581 & 1.147 & 4413 \\
54877.7 & 4.33 & 5830  & 1.216 & & 4.61 & 6016 & 1.296 & 4122 \\
54881.6 & 4.59 & 5544  & 1.402 & & 5.46 & 5207 & 1.667 & 3790 \\
54881.7 & 4.48 & 5655  & 1.371 & & 4.73 & 5979 & 1.448 & 3780 \\
54881.8 & 4.48 & 5694  & 1.373 & & 4.84 & 5960 & 1.486 & 3775 \\
54883.7 & 4.56 & 5538  & 1.455 & & 4.91 & 5701 & 1.570 & 3624 \\
54885.7 & 4.59 & 5508  & 1.528 & & 4.93 & 5791 & 1.641 & 3478 \\
54888.8 & 4.65 & 5497  & 1.644 & & 4.91 & 5928 & 1.739 & 3272 \\
54889.7 & 4.74 & 5292  & 1.707 & & 5.09 & 5554 & 1.837 & 3211 \\
54890.7 & 4.71 & 5351  & 1.732 & & 5.09 & 5382 & 1.873 & 3147 \\
54896.7 & 4.90 & 4992  & 2.019 & & 5.37 & 5198 & 2.212 & 2810 \\
54903.7 & 4.88 & 5019  & 2.281 & & 5.37 & 5195 & 2.512 & 2477 \\
\hline
\end{tabular}			  
\end{minipage}
\end{table*}

\subsection{Standardized Candle Method}\label{sec_scm}

The standardized candle method (SCM) was originally proposed by \citet{hamuy_scm} and has been refined several times \citep{hamuy_scm2,nugent, poznanski_scm, olivares_scm}. In this section we use multiple versions.

First we apply the version of \citet{poznanski_scm}. Using the measured brightness, expansion velocity and well-known distance of a sample of 34 SNe, they calibrated the equation
\begin{eqnarray}
\label{poz_eq}
\mathcal{M_I} - \alpha \cdot \log \left( {v_{Fe}(50\mathrm{d}) \over 5000} \right) +R_I((V-I) - (V-I)_0) - m_I =\\
= -5 \cdot \log(H_0D)\nonumber
\end{eqnarray}
where $\mathcal{M_I}=-1.615 \pm 0.08$, $\alpha=4.4 \pm 0.6$, $R_I=0.8 \pm 0.3$ and $(V-I)_0=0.53$~mag.  We measured the $I$-band magnitude of SN~2009N on day +50 as $m_I=15.440 \pm 0.015$~mag, the colour as $(V-I)=0.850 \pm 0.025$~mag, and $v_{\rm Fe}({\rm 50d})=2448 \pm 300$ km~s$^{-1}$.  We adopt the latest value of the Hubble constant, that was determined by the Planck collaboration: $H_0=67.3 \pm 1.2$~km~s$^{-1}$Mpc$^{-1}$ \citep{planck_cosmology}. The obtained distance is $D_{{\rm SCM},1}=21.01 \pm 2.86$~Mpc

\citet{maguire_scm} extended the technique by using NIR photometry and showed that in $JHK$ bands the scatter in the Hubble diagram is lower than in the optical. 
They used the same formula as \citet{poznanski_scm} (Eq.~\ref{poz_eq}). Their calibration in $J$ band -- which had the lowest scatter among the three bands -- led to the values of $\mathcal{M_J}=-2.532 \pm 0.250$~mag and $\alpha=6.33 \pm 1.20$. \citet{maguire_scm} opted to calculate with the value of $R_V=1.5$, obtained by \citet{poznanski_scm}, instead of using it as a fitting paramater. In the case of SN~2009N, $+50$ days after the explosion we measured $m_J=15.198 \pm 0.020$, $(V-J)=1.14 \pm 0.04$~mag and again used the velocity $v_{\rm Fe}({\rm 50d})=2448 \pm 300$ km~s$^{-1}$. The calculated distance is $D_{{\rm SCM},2}=20.82 \pm 4.35$~Mpc.

Both of the above methods rely on data taken on day $+50$ after explosion. However, in many cases the date of the explosion cannot be determined accurately. And even though the brightness does not change significantly around day +50, the velocity does; a few days uncertainty in the epoch of explosion can change the measured $v({\rm 50d})$ by $\pm~200-250$ km~s$^{-1}$. In the case of SN 2009N, $200$ km~s$^{-1}$ difference alone would change the distance by $\sim8$ per cent. A solution to this problem was proposed in \citet{olivares_scm}, who used an epoch calculated relative to the middle of the transition from the plateau to the tail. 

\citet{olivares_scm} examined the data of 37 nearby II-P SNe, and applied the same expression as \citet{hamuy_scm}:
\begin{equation}
m+\alpha \log(v_{Fe}/5000)-\beta(V-I)=5\log H_0D + zp,
\end{equation}
but using the magnitudes and velocities measured 30 days before the middle of the transition phase ($t_{PT}$). The values of $\alpha$, $\beta$ and $zp$ can be found in Table 6 of \citet{olivares_scm} for the filters $BVI$. In the case of SN~2009N we determined the middle of the transition by fitting the model function of \citet*{elmhamdi} to the light curve as $t_{PT}=109 \pm 2$ days after explosion. On day $t_{PT}-30$ we measured the values of $m_B=18.100 \pm 0.090$, $m_V=16.470 \pm 0.016$, $m_I=15.452 \pm 0.007$ magnitudes and $v_{\rm Fe}=1822 \pm 150\,\rm{km~s}^{-1}$. We determined the distance for all three bands, and calculated their weighted mean. This way we obtained a distance of $D_{{\rm SCM},3}=21.11 \pm 1.65$ Mpc.

The distances determined with the three versions of SCM agree quite well. We conclude their weighted mean, $D_{\rm SCM}=21.02 \pm 1.36$ Mpc ($\mu=31.61 \pm 0.14$~mag) as the SCM distance to SN 2009N. Note that SCM has multiple uncertainties. It requires external calibration using SNe with well-known distances and depends on the value of $H_0$. Adopting, for example, $H_0=73$ km~s$^{-1}$Mpc$^{-1}$ \citep{freedman_hubble} instead of $H_0=67.3$ km~s$^{-1}$Mpc$^{-1}$ would lower the SCM distance of SN~2009N by $\sim 8$ per cent.

\subsection{Average Distance}\label{sec_avedist}

The distances derived via EPM and SCM differ significantly, $D_{\rm EPM}$ being higher by $\sim10$ per cent than $D_{\rm SCM}$. \citet{olivares_scm} compared the SCM and EPM distances of SNe that were both in their sample and in the sample of \citet{jones2009}. They found that the EPM distances \citep[using the atmosphere models of ][]{D05} are systematically higher that those calculated via SCM. 

The distance of NGC 4487, the host galaxy of SN~2009N, was determined by \citet{tully_dist} as $D_{\rm TF}=19.9 \pm 4.0$~Mpc ($\mu =31.49 \pm 0.4$~mag) from the Tully-Fisher relation. This value is in agreement with our findings. 
On the other hand, the distance of the host galaxy derived from the redshift is significantly lower. After correcting for the Virgo infall a value of 
$15.3 \pm 0.9\,\rm{Mpc}$ can be obtained\footnote{HyperLeda, \url{http://leda.univ-lyon1.fr/}} (with $H_0=67.3\,{\rm km~s^{-1}Mpc^{-1}}$, as in the previous section). In the case of such nearby galaxies, peculiar motions can play an important role, and can explain the difference.  

By calculating the weighted average of the SCM, EPM and Tully-Fisher distances, and excluding the Hubble flow distance (Table \ref{ave-dist}), we determined our best estimate of the distance to NGC 4487 as $D=21.6 \pm 1.1$ Mpc ($\mu=31.67 \pm 0.11$).

\begin{table}
 \centering
 \begin{minipage}{84mm}
  \caption{Comparison of the distances obtained with different methods in this paper and in \citet{tully_dist}. Their weighted mean is also included.}
  \label{ave-dist}
  \begin{tabular}{@{}lccc@{}}
  \hline
  \hline
Method & Distance & Distance modulus & Ref. \\
& (Mpc) & (mag) \\
 \hline
Tully-Fisher & 19.9 (4.0) & 31.49 (0.44)  & \citet{tully_dist} \\
EPM & 23.3 (1.9) & 31.84 (0.18) & this paper \\
SCM & 21.0 (1.4) & 31.61 (0.14) & this paper\\
\\
average & 21.6 (1.1) & 31.67 (0.11) \\
\hline	
\end{tabular}			  
\end{minipage}
\end{table}

%####################### physical parameters ###################

\section{Physical parameters}\label{phys}

Based on our observations and the distance determined in the previous section, we infer some of the pysical parameters of the progenitor star and the explosion of SN 2009N.

The $^{56}$Ni mass produced during the explosion was estimated from the tail luminosity. Using eq.2. in \citet{hamuy2003}, we estimated the nickel mass to be $M_{\rm Ni}=0.020 \pm 0.004$~M$_{\sun}$.

With the same well-tested approach adopted for other observed CC-SNe \citep[e.g.~SNe 2007od, 2009bw, and 2009E; see][]{Inserra2011, inserra2012, pastorello_09E}, we evaluated
the main physical properties of the progenitor of SN 2009N at the explosion (i.e. the ejected 
mass, the progenitor radius, and the explosion energy) through the hydrodynamical modelling of 
the main observables (i.e. bolometric light curve, evolution of line velocities, and continuum 
temperature at the photosphere). 

According to this approach, a simultaneous $\chi^{2}$ fit of the above mentioned observables 
against model calculations was performed. Two codes were employed for the computation of the 
models: 1) the first one is a semi-analytic code where the energy balance equation is solved 
for a homologously expanding envelope of constant density \citep{zampieri2003, zampieri_sne_conf}; 
2) the second is a new general-relativistic, radiation-hydrodynamics Lagrangian code presented 
in \citet*{pumo2010} and \citet{pumo2011}, which is able to simulate the 
evolution of the physical properties of the CC SN ejecta and the behavior of the main observables 
from the breakout of the shock wave at the stellar surface up to the nebular stage. The distinctive 
features of this new code are: a) an accurate treatment of radiative transfer coupled to hydrodynamics, 
b) a fully implicit Lagrangian approach to the solution of the coupled non-linear finite difference 
system of general-relativistic, radiation-hydrodynamics equations, and c) a description of the evolution 
of ejected material which takes into account both the gravitational effects of the compact remnant 
and the heating effects linked to the decays of the radioactive isotopes synthesized during the SN 
explosion.

The semi-analytic code is used to carry out a preparatory study aimed at constraining the 
parameter space describing the SN progenitor at the explosion. The results of such study 
are exploited to guide the more realistic, but time consuming model calculations performed with 
the general-relativistic, radiation-hydrodynamics code.

We note that modelling with both codes is appropriate, since the emission of SN 2009N is dominated 
by the expanding ejecta. However, in performing the 
$\chi^{2}$ fit, we do not include the observational data taken at early phase (first $\sim$ $10-20$ days 
after explosion)
This is approximately the time needed for the early bolometric light curve to relax to the plateau \citep[see e.g.][]{tomasella_12A}. We do not attempt to model this phase because during it all the observables are significantly affected by emission from the outermost shell of the ejecta, which is not in homologous expansion \citep[cf.][]{pumo2011}. The structure, evolution and emission properties of this shell are not well reproduced in our simulations because at present we adopt an ``ad hoc'' initial density profile, not consistently derived from a post-explosion calculation. Future plans involve implementing more ``realistic'' density profiles in the simulations.

The shock breakout epoch, the bolometric luminosity, the photospheric velocity and the photospheric temperature were necessary to perform the comparison 
with model calculations (see Fig.~17). The agreement between our modelling and the observations is reasonably good apart from the early evolution of the photometric velocity. As mentioned before, the reason for this difference is due to the ``ad hoc'' initial density profile used in our simulations, which does not reproduce correctly the radial profile in the outermost shells of the ejecta formed after shock breakout. For this reason, we omit the early phase data from the fit. However, it should be noted that such omission does not affect the results significantly, the modelling  provides a reliable estimate of the main physical properties of the SN progenitor. 

Assuming a $^{56}$Ni mass of $0.020 \pm 0.004\,{\rm M_{\sun}}$, the 
the best fit model calculated with the general-relativistic,
radiation-hydrodynamics code returned values of total (kinetic plus thermal) energy of $\sim 0.48$ foe, initial radius of $\sim 2.0 \times 10^{13}$ cm ($\sim 287$~R$_{\sun}$), 
and envelope mass of $\sim 11.5\,{\rm M_{\sun}}$.

Adding the mass of a compact remnant ($\sim 1.5-2\,{\rm M_{\sun}}$)  to that of the ejected material, the estimated mass of the progenitor of SN 2009N at the explosion is $\sim 13-13.5\,{\rm M_{\sun}}$, which is consistent with the mass range of the red supergiant precursors of SNe II-P \citep[e.g.][]{smartt2009}. Examining the pre-explosion images of SN~2009N, the upper limit for the progenitor mass was determined as $M_{ZAMS}<16$ M$_{\sun}$ \citep[see][and the references therein]{maguire_nebusp}. The initial radius, however, is quite small for a RSG star, it is more consistent with that of a yellow supergiant.

\begin{figure}
\includegraphics[angle=270,width=84mm]{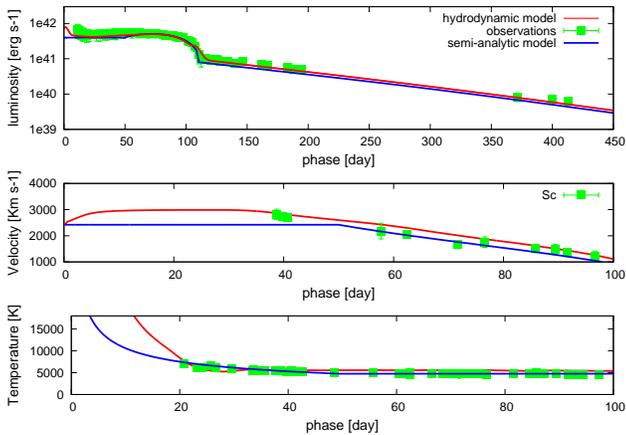}
\label{fighydro}
\caption{ Comparison of the evolution of the main observables of SN~2009N with the best-fit model computed with the general-relativistic, radiation-hydrodynamics code (total energy $\sim 0.48$~foe, initial radius $2.0 \times 10^{13}$~cm, envelope mass $\sim 11.5\,{\rm M_{\sun}}$). Top, middle, and bottom panels show the bolometric light curve, the photospheric velocity, and the photospheric temperature as a function of time. For the sake of completeness, the best-fit model computed with the semianalytic code (total energy $\sim 0.5$~foe, initial radius $\sim 1.1 \times 10^{13}$~cm, envelope mass $\sim 10\,{\rm M_{\sun}}$) is also reported. 
The photospheric velocities were estimated from the minima of the Sc\,{\sc ii} line profiles which are considered as good tracers of the photosphere velocity in Type II SNe. The values of the photospheric temperature taken at early phase (first 8 measurements corresponding to the first ∼ 15 days after the breakout of the
shock wave at the stellar surface) are not included in the fit (see text for details). As for the photospheric temperature, we use the black-body temperature derived from the black-body fits to the spectral continuum and to the $BVRIJH$ fluxes.}
\end{figure}

The estimated physical parameters of SN~2009N are in between those of the subluminous and normal SNe II-P. 
The ejected $^{56}$Ni mass ($0.020\,{\rm M_{\sun}}$)  is higher than that of the subluminous SNe 2005cs \citep[0.003-0.004~M$_{\sun}$,][]{pastorello05csII} and 2003Z \citep[0.0063~M$_{\sun}$,][]{utrobin03Z}, but lower than that of the normal II-P SN 1999em \citep[0.036~M$_{\sun}$,][]{utrobin99em}. Similarly, the explosion energy ($\sim 0.48$~foe) is in between the typical values of the subluminous (e.g. $0.2-0.4$~foe and $0.245$~foe for SNe 2005cs and 2003Z, respectively) and the normal SNe II-P \citep[e.g. $1.3$~foe for SN~1999em,][]{zampieri_sne_conf}. The initial radii and progenitor masses of SNe~2009N and 2003Z are alike, although the explosion energy of SN~2003Z was about half of that of SN~2009N \citep{utrobin03Z}.

The observational properties of SN 2009N are very similar to those of SN 2008in. The estimated $^{56}$Ni masses are also in good agreement. \citet{roy08in} used the semi-analytical formulae of \citet{litvinova} to estimate the physical parameters of the explosion and the progenitor of SN 2008in. They found that the progenitor was a compact star with $R_{ini} \approx 127\,{\rm R_{\sun}}$ and $M_{ej} \approx 17\,{\rm M_{\sun}}$. This radius is lower and the ejected mass is higher than those of SN~2009N. The explosion energies of the two SNe are similar. Using hydrodynamical modelling, \citet{utrobin_08in} also determined the parameters of SN~2008in. They obtained the ejecta mass of $13.6 \pm 1.9\,{\rm M_{\sun}} $ and the explosion energy of $0.51 \pm 0.34$ foe, which are close to the parameters inferred for SN~2009N. On the other hand, they estimated the radius of the progenitor of SN~2008in to be significantly larger than that of \citet{roy08in}, viz. $R_{\rm ini} = 570 \pm 100\,{\rm R_{\sun}}$. 

%########################################  summary ######################################3

\section{Summary}\label{summary}

In this paper we present ultraviolet, optical, and near-infrared photometry of SN 2009N together with optical and near-infrared spectra. The spectral and photometric evolution is similar to that of the intermediate luminosity SN II-P, 2008in \citep{roy08in}. The optical spectra show narrow features with low velocities, typical of subluminous SNe II-P. We examined the evolution of the strong Ba\,{\sc ii} $\lambda\lambda$5854 and 6497 lines, which are mainly detected in the spectra of subluminous SNe II-P. The bolometric luminosity during the plateau phase is in between those of the subluminous and normal SNe II-P. 

The NIR spectra of SN 2009N contain the usual features typical of SNe II-P, except for the appearance of a feature at $\sim 1.055\,{\rm\mu m}$ on day $+48$ after explosion. We also show that this feature is present in -- previously unpublished -- NIR spectra of SN~2008in. Creating {\sc synow} models of the spectra we found that this line is probably due to high-velocity He\,{\sc i} $\lambda 10830$, although we cannot rule out the identification as Si\,{\sc i} either. The presence of HV He\,{\sc i}, together with HV component of H$\alpha$, can be a sign of weak interaction of the ejecta with circumstellar material \citep{chugai_intera}.

We estimated the distance to SN~2009N using multiple versions of both the expanding photosphere method and the standardized candle method. As a result we determined the distance as $D=21.6 \pm 1.1$~Mpc ($\mu=31.67 \pm 0.11$).
The produced nickel mass is estimated to be $0.020 \pm 0.004$ M$_{\sun}$. Physical properties of the progenitor at the explosion were determined through hydrodynamical modelling. The total explosion energy ($\sim 0.48~{\rm foe}$) is in between the values typical of subluminous and normal SNe II-P. The presupernova mass ($\sim 13-13.5\,{\rm M_{\sun}}$) is consistent with that of RSG stars, while small estimated radius at the time of the explosion ($R_{\rm ini}\approx 287\,{\rm R_{\sun}}$) can point to a YSG star more than to a RSG.
The directly identified progenitors of normal SNe II-P, however, are all RSG stars \citep{smartt2009}, the only object for which the possibility of a YSG progenitor arose was SN~2008cn, a high-luminosity SNe II-P \citep{eliasrosa_2008cn}.

%###################################### acknowledgements #####################################

\section*{Acknowledgements}

K.T. acknowledges support by the Gemini-Conicyt project 32110024.
K.T., G.P., J.A., F.B., R.C., M.H. and F.F. acknowledge support from Millennium Center for Supernova Science (P10-064-F), with input from Fondo de Innovaci\' on para la Competitividad, del Ministerio de Economia, Fomento y Turismo de Chile. 
This project has been supported by the Hungarian OTKA grant NN~107637 and by the European Union together with the European Social Fund through the T\' AMOP 4.2.2/B-10/1-2010-0012 grant.
We acknowledge the TriGrid VL project and the INAF-Astronomical Observatory of Padua for the use
of computer facilities. M.L.P., A.P. and S.B. acknowledge support from the PRIN-INAF 2011 Transient
Universe: from ESO Large to PESSTO (P.I. S. Benetti).
N.E.R. acknowledges financial support by the MICINN grant
 AYA2011-24704/ESP, by the ESF EUROCORES Program EuroGENESIS (MINESCO grant
EUI2009-04170), and from the European Union Seventh Framework Programme (FP7/2007-2013) under grant agreement n. 267251.
F.B., J. A. and F.F. acknowledges support from CONICYT through FONDECYT grants 3120227, 3110142 and 3110042, respectively. F.F. acknowledges partial support from Comite Mixto ESO-GOBIERNO DE CHILE.
R.C. acknowledges support by CONICYT through “Programa Nacional de Becas
de Postgrado”
grant D-2108082, and by the Yale-Chile fellowship in astrophysics.
G.L. is supported by the Swedish Research Council through grant No. 623-2011-7117
M.S. gratefully acknowledge generous support provided by the Danish Agency for Science and Technology and Innovation  
realized through a Sapere Aude Level 2 grant.

This work is partially based on observations made with ESO Telescopes at the
La Silla and Paranal Observatories under programme IDs 084.D-0261 and 082.A-0526, and 
on observations of the European supernova
collaboration involved in the ESO-NTT large programme 184.D-1140 led by
Stefano Benetti.
This research is based in part on observations made with the Liverpool
Telescope operated on the island of La Palma by Liverpool John Moores
University in the Spanish Observatorio del Roque de los
Muchachos of the Instituto de Astrofisica de Canarias with financial
support from the UK Science and Technology Facilities Council; the
Nordic Optical Telescope, operated on the island of La Palma jointly
by Denmark, Finland, Iceland, Norway, and Sweden, in the Spanish
Observatorio del Roque de los Muchachos of the Instituto de
Astrofisica de Canarias; the SMARTS Consortium 1.3~m
telescope and the Prompt Telescopes located at Cerro Tololo Inter-American Observatory (CTIO),
Chile; the 1.5~m telescope located at Palomar Observatory, USA; the 2.2 m
telescope
of the Calar Alto Observatory (Sierra de Los Filabres, Spain); the Southern Astrophysical Research (SOAR) telescope, which is a joint project of the Minist\'{e}rio da Ci\^{e}ncia, Tecnologia, e Inova\c{c}\~{a}o (MCTI) da Rep\'{u}blica Federativa do Brasil, the U.S. National Optical Astronomy Observatory (NOAO), the University of North Carolina at Chapel Hill (UNC), and Michigan State University (MSU);  the 2.5 m du Pont Telescope and the 6.5 m Magellan Telescopes located at Las Campanas Observatory, Chile.
We are grateful to the staffs at these observatories for their excellent
assistance with the observations.

This research has made use of the NASA/IPAC Extragalactic Database, the HyperLeda database, NASA's Astrophysics Data System. The availability of these services is gratefully acknowledged.

We thank Rupak Roy for sending us the optical spectra of SN~2008in and Morgan Fraser for the NIR spectra of SN~2009md. We also thank the referee, V.~P. Utrobin for the thorough review of the paper.

%##########################################x  biblio   ############################################
\bibliographystyle{mn2e}
\bibliography{references}

%###########################################  appendix  ############################################3
\appendix

\section{Near infrared spectra of SN 2008in}\label{app_a}

\begin{figure}
\includegraphics[width=84mm]{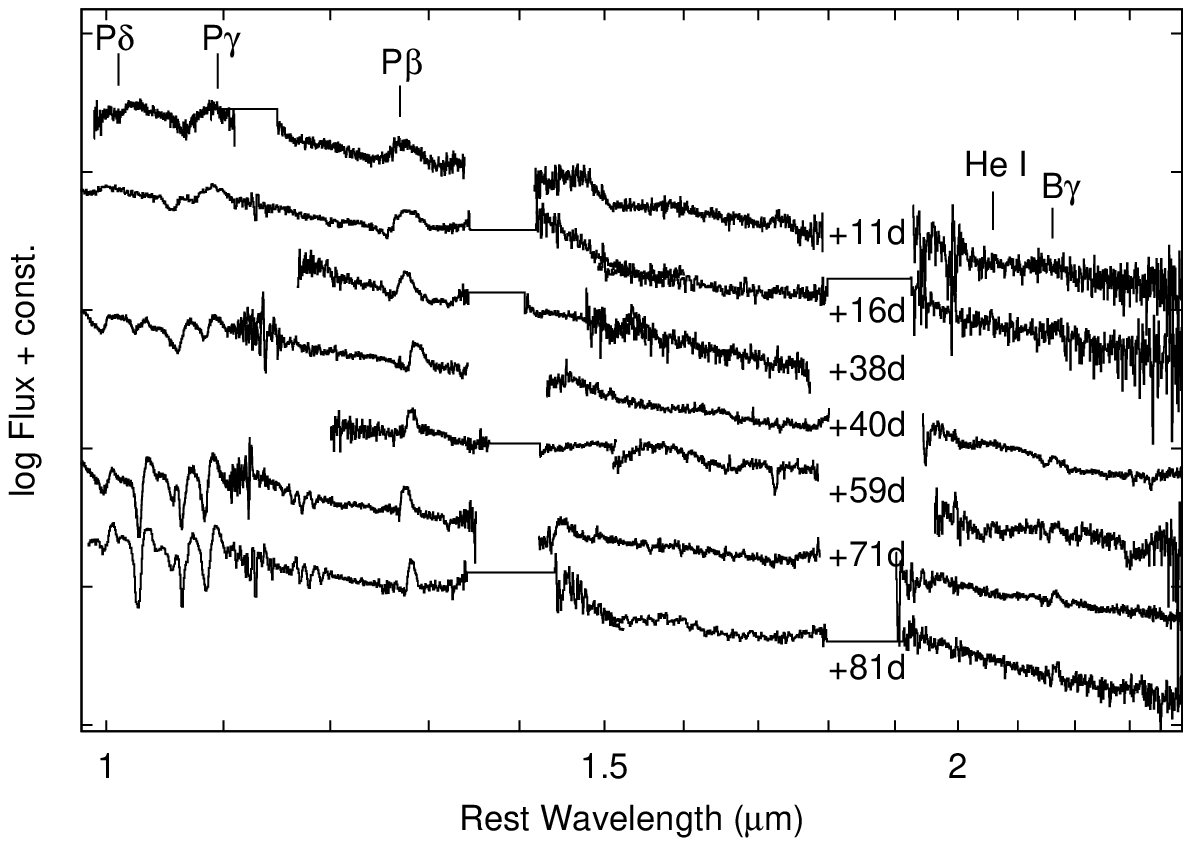}
\caption{Near infrared spectra of SN 2008in (see also Fig. \ref{nir_seq}). The phases are relative to the estimated date of explosion, $t_0=2454825.6$~JD \citep{roy08in}.}
\label{08in_nir_seq1}
\end{figure}

In this paper we present 7, previously unpublished near infrared spectra of SN 2008in taken with VLT/ISAAC, NTT/SOFI, and SOAR/OSIRIS telescopes. The summary of the observations can be found in Table \ref{logofnirsp_08in}. The reduction process was the same as for SN 2009N (see Sec. \ref{spectroscopy}). The NIR spectra of SN~2008in are shown in Fig. \ref{08in_nir_seq1}.

 \begin{table*}
 \centering
 \begin{minipage}{\textwidth}
  \caption{Summary of the near infrared spectroscopic observations of SN 2008in.}
  \label{logofnirsp_08in}
  \begin{tabular}{@{}cccllll@{}}
  \hline
  \hline
Date & JD & Phase$^a$ & Instrument &  Wavelength range &  Resolution \\
     & 2400000+ & days & & $\mu$m &  \\
  \hline 
 05/01/2009 & 54836.8 & 11.2 & VLT + ISAAC (SWS1-LR) &  0.98-2.5 &  500\\
 10/01/2009 & 54841.8 & 16.2 & NTT + SOFI (GB,GR)  & 0.95-1.64, 1.53-2.52 & 1000 \\
 01/02/2009 & 54863.8 & 38.2 & SOAR + OSIRIS (LR) & 1.0-2.58 & 1200  \\ 
 03/02/2009 & 54865.9 & 40.3 & VLT + ISAAC (SWS1-LR) & 0.98-2.5 & 500 \\
 22/02/2009 & 54884.7 & 59.1 & SOAR+OSIRIS (LR) & 1.0--2.58 & 1200 \\
 05/03/2009 & 54895.7 & 70.1 & VLT + ISAAC (SWS1-LR) & 0.98-2.5 & 500 \\
 17/03/2009 & 54907.2 & 81.6 & NTT + SOFI (GB,GR) & 0.95-1.64, 1.53-2.52 & 1000 \\
\hline				  
\end{tabular}			  
 \begin{tablenotes}
       \item[a]{$^a$ relative to $t_0=2454825.6$~JD \citep{roy08in}.}
     \end{tablenotes}
\end{minipage}
\end{table*}

\section{Photometric tables of SN 2009N}

\begin{table*}
 \centering
 \begin{minipage}{170mm}
  \caption{$BVRI$ photometry of SN 2009N.}
  \label{lc_table}
  \begin{tabular}{@{}ccccccc@{}}
  \hline
  \hline
  date  & JD$-2400000$ & $B$ & $V$ & $R$ & $I$ & Telescope \\
 \hline
27/01/2009 & 54858.9 & 16.585 (0.020) & 16.274 (0.027) & 15.931 (0.020) & 15.767 (0.028) & P60 \\
28/01/2009 & 54859.7 & 16.474 (0.005) & 16.270 (0.004) & 15.973 (0.003) & 15.815 (0.005) & LT  \\
28/01/2009 & 54859.8 & 16.614 (0.030) & 16.299 (0.016) & 16.019 (0.014) & 15.832 (0.014) & PROMPT \\
29/01/2009 & 54860.8 & 16.559 (0.034) & 16.331 (0.020) & 16.033 (0.015) &                & PROMPT \\
29/01/2009 & 54860.9 & 16.477 (0.069) & 16.247 (0.026) & 15.973 (0.028) & 15.769 (0.029) & P60 \\
31/01/2009 & 54862.8 & 16.590 (0.028) & 16.337 (0.009) & 15.948 (0.016) & 15.751 (0.011) & P60 \\
31/01/2009 & 54862.8 & 16.633 (0.023) & 16.312 (0.019) & 16.006 (0.016) & 15.791 (0.015) & PROMPT \\
01/02/2009 & 54863.8 & 16.654 (0.029) & 16.290 (0.021) & 15.998 (0.017) & 15.764 (0.017) & PROMPT \\
02/02/2009 & 54864.8 & 16.694 (0.021) & 16.341 (0.014) & 15.955 (0.019) & 15.664 (0.016) & P60 \\
04/02/2009 & 54866.8 & 16.831 (0.025) & 16.314 (0.019) & 15.978 (0.016) & 15.712 (0.016) & PROMPT \\
06/02/2009 & 54868.8 & 16.957 (0.031) & 16.347 (0.018) & 15.994 (0.016) & 15.709 (0.015) & PROMPT \\
09/02/2009 & 54871.8 & 		      & 16.369 (0.026) & 15.991 (0.015) &		 & PROMPT \\
12/02/2009 & 54874.7 & 		      & 16.392 (0.021) & 16.004 (0.019) & 15.676 (0.018) & PROMPT \\
15/02/2009 & 54877.7 & 17.368 (0.026) & 16.393 (0.019) & 16.016 (0.015) & 15.671 (0.017) & PROMPT \\
17/02/2009 & 54879.6 &		      &		       & 15.937 (0.023) &		 & THO  \\
17/02/2009 & 54879.7 & 17.373 (0.034) & 16.405 (0.018) & 16.005 (0.018) & 		 & PROMPT \\
19/02/2009 & 54881.6 & 17.407 (0.012) & 16.344 (0.020) & 15.889 (0.022) & 15.544 (0.033) & CA  \\
19/02/2009 & 54881.7 & 17.394 (0.036) & 16.392 (0.018) & 15.957 (0.016) & 15.605 (0.015) & PROMPT \\
19/02/2009 & 54881.8 & 17.343 (0.026) &		       & 15.936 (0.011) &		 & P60 \\
19/02/2009 & 54882.0 & 	              & 16.382 (0.010) &  	        & 15.564 (0.013) & P60 \\
20/02/2009 & 54882.6 & 	              &		       & 15.975 (0.023) &		 & CA \\
21/02/2009 & 54883.7 & 17.446 (0.029) & 16.374 (0.019) & 15.964 (0.015) & 15.569 (0.015) & PROMPT \\
23/02/2009 & 54885.7 & 17.432 (0.038) & 16.382 (0.017) & 15.942 (0.015) & 15.541 (0.015) & PROMPT \\
25/02/2009 & 54888.5 &		      &		       & 15.905 (0.022) &		 & THO \\
26/02/2009 & 54888.8 & 17.406 (0.058) & 16.352 (0.021) & 15.894 (0.031) & 15.527 (0.020) & P60 \\
27/02/2009 & 54889.7 & 17.547 (0.037) & 16.360 (0.019) & 15.914 (0.014) & 15.506 (0.014) & PROMPT \\
28/02/2009 & 54890.7 & 17.535 (0.032) & 16.337 (0.033) & 15.949 (0.040) & 15.515 (0.011) & P60 \\
06/03/2009 & 54896.7 & 17.761 (0.037) & 16.372 (0.018) & 15.880 (0.014) & 15.455 (0.014) & PROMPT \\
13/03/2009 & 54903.7 & 17.724 (0.058) & 16.390 (0.028) & 15.892 (0.020) & 15.464 (0.026) & PROMPT \\
15/03/2009 & 54905.7 & 17.760 (0.034) & 16.413 (0.019) & 15.858 (0.015) &		 & PROMPT \\
17/03/2009 & 54908.0 & 17.752 (0.015) &		       & 15.857 (0.028) &		 & P60 \\
18/03/2009 & 54908.6 & 17.914 (0.039) & 16.416 (0.019) & 15.859 (0.015) & 15.402 (0.014) & PROMPT \\
19/03/2009 & 54909.7 & 17.928 (0.040) & 16.414 (0.017) & 15.861 (0.014) & 15.403 (0.014) & PROMPT \\
19/03/2009 & 54910.5 &		      &		       & 15.823 (0.066) &		 & THO \\
19/03/2009 & 54910.5 & 17.842 (0.010) &	16.545 (0.014) & 15.853 (0.011) & 15.364 (0.016) & CA \\ 
20/03/2009 & 54910.5 & 17.747 (0.014) &	16.470 (0.008) & 15.878 (0.014) & 15.417 (0.007) & NOT  \\
21/03/2009 & 54912.5 &		      &		       & 15.850 (0.018) &		 & THO \\
22/03/2009 & 54912.7 &		      & 16.432 (0.019) & 15.857 (0.015) & 15.405 (0.014) & PROMPT \\
24/03/2009 & 54914.6 & 17.976 (0.044) & 16.434 (0.018) & 15.855 (0.014) & 15.421 (0.014) & PROMPT \\
25/03/2009 & 54915.5 &		      &		       & 15.854 (0.022) &		 & THO \\
26/03/2009 & 54916.6 & 17.907 (0.051) & 16.447 (0.018) & 15.880 (0.014) & 15.404 (0.014) & PROMPT \\
26/03/2009 & 54916.9 & 17.863 (0.040) & 16.505 (0.016) & 15.858 (0.007) & 15.344 (0.008) & P60 \\
27/03/2009 & 54917.9 & 17.850 (0.067) & 16.496 (0.030) & 15.928 (0.021) & 15.403 (0.014) & P60 \\
28/03/2009 & 54918.6 & 18.003 (0.055) & 16.483 (0.018) & 15.873 (0.015) & 15.406 (0.014) & PROMPT \\
28/03/2009 & 54919.0 & 17.919 (0.048) & 16.527 (0.013) & 15.887 (0.011) & 15.400 (0.008) & P60 \\
30/03/2009 & 54921.4 & 17.956 (0.025) & 16.501 (0.006) & 15.833 (0.004) & 15.467 (0.005) & LT  \\
31/03/2009 & 54921.0 & 17.901 (0.030) & 16.528 (0.011) & 15.873 (0.009) & 15.434 (0.008) & P60 \\
01/04/2009 & 54922.6 & 17.922 (0.050) & 16.503 (0.017) & 15.888 (0.014) & 15.422 (0.014) & PROMPT \\
01/04/2009 & 54923.4 & 18.045 (0.013) & 16.477 (0.007) & 15.857 (0.005) & 15.460 (0.004) & LT  \\
01/04/2009 & 54923.5 &		      &		       & 15.830 (0.024) &		 & THO \\
05/04/2009 & 54927.4 & 	              & 16.527 (0.016) &		& 15.452 (0.007) & LT  \\
08/04/2009 & 54929.6 & 18.210 (0.069) & 16.498 (0.028) & 15.947 (0.018) & 15.460 (0.018) & PROMPT \\
11/04/2009 & 54932.6 & 18.139 (0.053) & 16.598 (0.024) & 15.996 (0.016) & 15.535 (0.015) & PROMPT \\
11/04/2009 & 54933.5 & 18.210 (0.093) &	16.622 (0.017) & 16.004 (0.010) & 15.570 (0.009) & NOT \\
11/04/2009 & 54933.5 & 18.096 (0.058) &	16.686 (0.014) & 16.011 (0.012) & 15.556 (0.010) & CA \\
12/04/2009 & 54933.5 & 18.133 (0.030) &	16.573 (0.013) & 15.925 (0.006) & 15.557 (0.006) & LT \\
13/04/2009 & 54934.7 & 18.115 (0.022) &	16.605 (0.015) & 15.999 (0.007) & 15.523 (0.020) & P60 \\
13/04/2009 & 54935.4 & 18.190 (0.015) &	16.628 (0.006) & 15.954 (0.009) & 15.565 (0.005) & LT  \\
15/04/2009 & 54937.4 &		      &		       & 15.989 (0.020) &		 & THO \\
16/04/2009 & 54937.6 & 18.244 (0.030) &	16.659 (0.007) & 15.968 (0.008) & 15.617 (0.008) & LT  \\
18/04/2009 & 54939.7 & 18.306 (0.038) & 16.708 (0.019) & 16.081 (0.015) & 15.601 (0.013) & PROMPT \\
\hline				  
\end{tabular}			  
\end{minipage}
\end{table*}

\begin{table*}
 \centering
 \begin{minipage}{170mm}
  \contcaption{}
  \label{}
  \begin{tabular}{@{}ccccccc@{}}
  \hline
  \hline
  date  & JD$-2400000$ & $B$ & $V$ & $R$ & $I$ & Telescope \\
 \hline
19/04/2009 & 54941.5 & 18.327 (0.035) & 16.763 (0.009) & 16.059 (0.011) & 15.672 (0.009) & LT  \\
21/04/2009 & 54942.6 & 18.377 (0.056) & 16.796 (0.018) & 16.133 (0.014) & 15.669 (0.015) & PROMPT \\
23/04/2009 & 54945.4 & 18.471 (0.048) & 16.886 (0.021) & 16.166 (0.008) & 15.774 (0.011) & LT  \\
23/04/2009 & 54945.4 &		      &		       & 16.191 (0.037) &		 & THO \\
24/04/2009 & 54945.6 & 18.596 (0.052) & 16.841 (0.027) & 		&		 & PROMPT \\
25/04/2009 & 54946.6 &		      &		       & 16.280 (0.015) & 15.762 (0.014) & PROMPT \\
25/04/2009 & 54947.4 &		      &		       & 16.210 (0.030) &		 & THO \\
27/04/2009 & 54948.8 & 18.545 (0.038) &		       & 16.340 (0.005) &		 & P60 \\
28/04/2009 & 54949.7 & 18.590 (0.043) & 17.090 (0.014) & 16.314 (0.009) & 15.801 (0.012) & P60 \\ 
29/04/2009 & 54950.6 & 18.712 (0.059) & 17.040 (0.022) & 16.374 (0.017) & 15.906 (0.016) & PROMPT \\
29/04/2009 & 54951.4 & 18.757 (0.021) & 17.134 (0.009) & 16.387 (0.006) & 15.918 (0.006) & LT  \\
30/04/2009 & 54952.4 &		      &		       & 16.445 (0.037) &		 & THO \\
01/05/2009 & 54952.6 & 18.726 (0.067) & 17.183 (0.022) & 16.456 (0.015) & 15.995 (0.016) & PROMPT \\
03/05/2009 & 54955.4 & 19.114 (0.025) & 17.483 (0.010) & 16.701 (0.008) & 16.215 (0.007) & LT  \\
04/05/2009 & 54955.6 & 		      & 17.469 (0.032) & 16.730 (0.018) & 16.185 (0.019) & PROMPT \\
07/05/2009 & 54958.5 & 19.075 (0.276) & 18.343 (0.155) & 17.261 (0.068) & 16.864 (0.066) & PROMPT \\
08/05/2009 & 54960.4 & 20.201 (0.048) & 18.504 (0.018) & 17.690 (0.020) & 17.069 (0.008) & NOT \\
09/05/2009 & 54960.6 &		      & 18.388 (0.082) & 17.558 (0.038) & 16.967 (0.035) & PROMPT \\
10/05/2009 & 54961.7 & 19.474 (0.143) &		       & 17.580 (0.038) & 16.993 (0.035) & PROMPT \\
11/05/2009 & 54962.6 &		      &		       & 17.575 (0.033) & 17.095 (0.029) & PROMPT \\
11/05/2009 & 54963.4 & 20.438 (0.058) & 18.808 (0.022) & 17.693 (0.014) & 17.158 (0.010) & LT  \\
12/05/2009 & 54963.6 & 20.126 (0.193) & 18.575 (0.076) & 17.633 (0.031) & 17.108 (0.029) & PROMPT \\
12/05/2009 & 54963.7 & 20.347 (0.451) & 18.739 (0.099) & 17.677 (0.060) & 17.123 (0.015) & P60 \\
13/05/2009 & 54964.6 & 20.172 (0.155) & 18.641 (0.065) & 17.643 (0.031) & 17.142 (0.022) & PROMPT \\
13/05/2009 & 54964.7 & 20.374 (0.299) & 18.774 (0.083) & 17.704 (0.041) & 17.060 (0.010) & P60 \\
16/05/2009 & 54968.4 & 20.509 (0.087) & 18.796 (0.054) & 17.785 (0.017) & 17.222 (0.018) & LT  \\
20/05/2009 & 54971.6 & 		      & 18.890 (0.130) & 17.707 (0.039) & 17.184 (0.050) & PROMPT \\
22/05/2009 & 54973.6 &		      & 18.665 (0.053) & 17.735 (0.026) & 17.172 (0.021) & PROMPT \\
22/05/2009 & 54974.5 & 20.576 (0.304) & 18.941 (0.032) & 17.792 (0.018) & 17.236 (0.023) & LT  \\
24/05/2009 & 54975.7 & 20.395 (0.190) & 18.919 (0.068) & 17.867 (0.028) & 17.277 (0.013) & P60 \\
25/05/2009 & 54977.4 & 20.499 (0.188) & 18.941 (0.049) & 17.830 (0.023) & 17.304 (0.017) & LT  \\
27/05/2009 & 54978.7 &	  	      & 18.979 (0.025) & 17.882 (0.017) & 17.248 (0.035) & P60 \\
30/05/2009 & 54982.4 & 20.527 (0.073) & 18.975 (0.041) & 17.860 (0.012) & 17.375 (0.018) & LT \\
05/06/2009 & 54987.7 &		      & 18.925 (0.125) & 17.970 (0.038) & 17.350 (0.056) & P60 \\
12/07/2009 & 54994.6 & 		      & 18.864 (0.067) & 17.884 (0.022) & 17.359 (0.021) & PROMPT \\
27/06/2009 & 55009.7 &		      &	19.096 (0.033) & 18.076 (0.015) & 17.505 (0.045) & P60 \\
02/07/2009 & 55014.7 &		      &	19.120 (0.091) & 18.127 (0.038) & 17.621 (0.023) & P60 \\
18/07/2009 & 55031.5 &		      &	19.344 (0.040) & 18.306 (0.026) & 17.713 (0.070) & CTIO \\
23/07/2009 & 55036.5 & 	     	      & 19.364 (0.038) & 18.432 (0.020) & 17.758 (0.018) & CTIO \\
29/07/2009 & 55042.5 &		      &	19.423 (0.032) & 18.451 (0.017) & 17.796 (0.023) & CTIO \\
23/01/2010 & 55219.8 &		      &	21.035 (0.075) & 20.419 (0.068) & 19.890 (0.070) & NTT \\
20/02/2010 & 55247.7 &		      &	21.160 (0.049) & 20.571 (0.044) & 20.017 (0.122) & NTT \\
05/03/2010 & 55260.8 &		      &	21.256 (0.056) & 20.670 (0.052) & 20.142 (0.044) & NTT \\
\hline				  
\end{tabular}			  
 \begin{tablenotes}
       \item[a]{P60 - Palomar 1.5-m telescope;  LT - 2.0-m Liverpool Telescope + RatCAM; PROMPT -  Panchromatic Robotic Optical Monitoring and Polarimetry Telescopes; THO - 14" Telescope at Taurus Hill Observatory + ST-8XME; CA - 2.2m Calar Alto + CAFOS; NOT - 2.6-m Nordic Optical Telescope + ALFOSC; CTIO - CTIO 1.3 m SMARTS telescope + ANDICAM; NTT - 3.6-m New Technology Telescope + EFOSC2}
     \end{tablenotes}
\end{minipage}
\end{table*}

\begin{table*}
 \centering
 \begin{minipage}{170mm}
  \caption{$g'r'i'z'$ photometry of SN 2009N taken with the $PROMPT$ telescopes.}
  \label{lc_table_griz}
  \begin{tabular}{@{}cccccc@{}}
  \hline
  \hline
  date  & JD$-2400000$ & $g'$ & $r'$ & $i'$ & $z'$  \\
 \hline
31/01/2009 &   54862.8 & 16.450 (0.015) & 16.143 (0.034) & 16.102 (0.017) & 16.067 (0.024) \\
01/02/2009 &   54863.8 & 16.501 (0.017) & 16.134 (0.035) & 16.050 (0.017) & 15.976 (0.021) \\
04/02/2009 &   54866.8 & 16.533 (0.018) & 16.132 (0.035) & 16.053 (0.019) & 15.931 (0.019) \\
06/02/2009 &   54868.8 & 16.627 (0.018) & 16.138 (0.034) & 16.050 (0.017) & 15.948 (0.020) \\
08/02/2009 &   54870.8 & 16.672 (0.017) & 16.166 (0.035) & 16.039 (0.018) & 	           \\
12/02/2009 &   54874.8 & 	        & 16.168 (0.035) & 16.026 (0.020) & 15.894 (0.026) \\
11/02/2009 &   54874.8 & 	        & 16.168 (0.035) & 16.026 (0.020) & 15.894 (0.026) \\
15/02/2009 &   54877.8 & 16.780 (0.015) & 16.171 (0.034) & 16.017 (0.017) & 15.824 (0.021) \\
17/02/2009 &   54879.7 & 16.824 (0.016) & 16.151 (0.035) & 	          & 	           \\
19/02/2009 &   54881.7 & 16.808 (0.017) & 16.148 (0.034) & 15.955 (0.017) & 15.834 (0.022) \\
21/02/2009 &   54883.7 & 16.835 (0.018) & 16.134 (0.034) & 15.932 (0.016) & 15.783 (0.020) \\
23/02/2009 &   54885.7 & 16.830 (0.018) & 16.132 (0.034) & 15.904 (0.017) & 15.740 (0.020) \\
27/02/2009 &   54889.7 & 16.847 (0.018) & 16.099 (0.034) & 15.861 (0.016) & 15.706 (0.019) \\
06/03/2009 &   54896.7 & 16.901 (0.016) & 16.086 (0.034) & 15.814 (0.015) & 15.664 (0.018) \\
13/03/2009 &   54903.7 & 16.949 (0.022) & 16.056 (0.039) & 15.730 (0.024) & 15.597 (0.042) \\
15/03/2009 &   54905.7 & 	        & 16.061 (0.034) & 15.763 (0.016) & 15.615 (0.021) \\
18/03/2009 &   54908.7 & 17.016 (0.015) & 16.059 (0.034) & 15.755 (0.016) & 15.586 (0.022) \\
19/03/2009 &   54909.7 & 17.017 (0.017) & 16.059 (0.034) & 15.773 (0.015) & 15.611 (0.018) \\
22/03/2009 &   54912.6 & 	        & 16.068 (0.035) & 15.760 (0.016) & 15.613 (0.020) \\
24/03/2009 &   54914.6 & 17.079 (0.017) & 16.074 (0.034) & 15.778 (0.016) & 15.634 (0.019) \\
26/03/2009 &   54916.6 & 17.101 (0.019) & 16.070 (0.035) & 15.774 (0.016) & 15.615 (0.020) \\
28/03/2009 &   54918.7 & 17.104 (0.019) & 16.084 (0.034) & 15.787 (0.016) & 15.606 (0.021) \\
01/04/2009 &   54922.6 & 17.115 (0.020) & 16.075 (0.034) & 15.780 (0.017) & 15.645 (0.020) \\
08/04/2009 &   54929.6 & 17.192 (0.027) & 16.176 (0.037) & 15.871 (0.022) & 15.743 (0.026) \\
11/04/2009 &   54932.6 & 17.242 (0.023) & 16.206 (0.036) & 15.848 (0.017) & 15.772 (0.022) \\
18/04/2009 &   54939.7 & 17.376 (0.018) & 16.259 (0.034) & 15.946 (0.016) & 15.800 (0.019) \\
21/04/2009 &   54942.6 & 17.412 (0.023) & 16.340 (0.035) & 16.019 (0.016) & 15.851 (0.021) \\
24/04/2009 &   54945.6 & 17.561 (0.020) & 16.413 (0.040) & 16.095 (0.023) & 	           \\
29/04/2009 &   54950.6 & 17.686 (0.040) & 16.596 (0.034) & 16.251 (0.019) & 16.076 (0.021) \\
01/05/2009 &   54952.6 & 17.866 (0.023) & 16.649 (0.034) & 16.334 (0.018) & 16.165 (0.023) \\
04/05/2009 &   54955.6 & 18.203 (0.030) & 16.917 (0.036) & 16.605 (0.021) & 16.432 (0.029) \\
07/05/2009 &   54958.6 & 18.540 (0.204) & 17.662 (0.090) & 	          & 	           \\
09/05/2009 &   54960.6 &  	        & 17.752 (0.050) & 17.395 (0.035) & 17.106 (0.048) \\
10/05/2009 &   54961.6 & 19.008 (0.061) & 17.823 (0.046) & 17.463 (0.029) & 17.193 (0.050) \\
11/05/2009 &   54962.6 & 	        & 17.900 (0.043) & 	          & 17.097 (0.051) \\
12/05/2009 &   54963.5 & 19.049 (0.043) & 17.926 (0.041) & 17.473 (0.029) & 17.295 (0.047) \\
23/05/2009 &   54974.6 & 19.387 (0.066) & 17.982 (0.041) & 17.634 (0.026) & 17.271 (0.042) \\
13/05/2009 &   54995.5 & 	        & 18.124 (0.040) & 17.900 (0.025) & 17.500 (0.035) \\
\hline				  
\end{tabular}			   
\end{minipage}
\end{table*}

\begin{table*}
 \centering
 \begin{minipage}{170mm}
  \caption{$BVRI$ magnitudes of the local sequence stars.}
  \label{std}
  \begin{tabular}{@{}ccccccc@{}}
  \hline
  \hline
  Star & $\alpha_{\rm J2000}$ & $\delta_{\rm J2000}$ & $B$ & $V$ & $R$ & $I$ \\
 \hline
 1 &  12:31:09.006 & $-$08:04:35.29 & 20.612 (0.238) & 19.225 (0.090) & 18.659 (0.031) & 17.979 (0.043) \\
 2 &  12:31:10.321 & $-$08:03:56.69 & 19.115 (0.076) & 18.158 (0.066) & 17.607 (0.040) & 17.026 (0.043) \\ 
 3 &  12:31:15.903 & $-$08:02:14.07 & 17.276 (0.101) & 16.665 (0.051) & 16.316 (0.038) & 15.893 (0.035) \\ 
 4 &  12:31:15.307 & $-$08:01:58.17 & 15.819 (0.088) & 15.118 (0.045) & 14.734 (0.031) & 14.316 (0.036) \\ 
 5 &  12:31:08.100 & $-$08:02:01.29 & 17.016 (0.074) & 16.252 (0.051) & 15.848 (0.035) & 15.411 (0.044) \\ 
 6 &  12:31:03.358 & $-$08:02:13.81 & 13.821 (0.078) & 12.920 (0.050) & 12.457 (0.029) & 11.969 (0.040) \\ 
 7 &  12:30:57.996 & $-$08:01:32.58 & 16.384 (0.064) & 15.624 (0.049) & 15.215 (0.026) & 14.756 (0.036) \\ 
 8 &  12:30:57.482 & $-$08:05:28.19 & 18.745 (0.163) & 17.043 (0.051) & 16.070 (0.030) & 14.836 (0.053) \\ 
 9 &  12:31:01.452 & $-$08:06:14.02 & 15.125 (0.061) & 14.335 (0.049) & 13.941 (0.035) & 13.545 (0.028) \\ 
10 &  12:31:07.750 & $-$08:06:39.12 & 17.154 (0.053) & 16.370 (0.048) & 15.959 (0.036) & 15.514 (0.032) \\ 
11 &  12:31:19.407 & $-$08:07:03.51 & 18.151 (0.983) & 16.936 (0.078) & 16.721 (0.038) & 16.380 (0.037) \\ 
12 &  12:31:22.057 & $-$08:06:19.63 & 18.379 (0.090) & 16.699 (0.045) & 15.753 (0.025) & 14.481 (0.049) \\ 
13 &  12:31:25.355 & $-$08:06:37.04 & 20.780 (0.010) & 15.838 (0.056) & 15.508 (0.042) & 15.086 (0.034) \\ 
14 &  12:31:24.593 & $-$08:03:59.05 & 20.780 (0.010) & 13.829 (0.043) & 13.537 (0.032) & 13.183 (0.032) \\ 
15 &  12:31:14.787 & $-$08:03:38.06 & 13.298 (0.073) & 12.607 (0.049) & 12.250 (0.038) & 11.884 (0.029) \\ 
16 &  12:31:27.718 & $-$08:02:34.53 & 20.780 (0.010) & 12.659 (0.489) & 11.671 (0.083) & 11.117 (0.025) \\ 
17 &  12:31:28.057 & $-$08:01:05.10 & 20.780 (0.010) & 18.141 (2.299) & 19.332 (6.133) & 15.229 (0.013) \\ 
18 &  12:31:21.260 & $-$08:00:22.14 & 17.736 (0.122) & 17.140 (0.060) & 16.809 (0.031) & 16.444 (0.045) \\ 
19 &  12:31:21.166 & $-$08:00:11.70 & 18.639 (0.096) & 17.190 (0.056) & 16.290 (0.035) & 15.166 (0.011) \\ 
20 &  12:31:17.638 & $-$07:59:52.25 & 16.493 (0.094) & 15.954 (0.042) & 15.641 (0.027) & 15.269 (0.032) \\ 
21 &  12:31:12.089 & $-$08:00:13.41 & 15.418 (0.107) & 14.828 (0.047) & 14.509 (0.028) & 14.143 (0.036) \\ 
22 &  12:31:06.949 & $-$07:59:24.83 & 17.077 (0.084) & 16.519 (0.045) & 16.213 (0.034) & 15.860 (0.030) \\ 
23 &  12:31:05.295 & $-$07:59:05.75 & 17.953 (0.139) & 16.986 (0.039) & 16.440 (0.019) & 15.882 (0.029) \\ 
24 &  12:31:05.463 & $-$07:58:54.88 & 18.738 (0.030) & 17.192 (0.054) & 16.344 (0.016) & 15.497 (0.045) \\ 
\hline				  
\end{tabular}			  
\end{minipage}
\end{table*}

\begin{table*}
 \centering
 \begin{minipage}{170mm}
  \caption{$g'r'i'z'$ magnitudes of the local sequence stars.}
  \label{std_sloan}
  \begin{tabular}{@{}ccccccc@{}}
  \hline
  \hline
  Star & $\alpha_{\rm J2000}$ & $\delta_{\rm J2000}$ & $g'$ & $r'$ & $i'$ & $z'$ \\
 \hline

 1 & 12:31:09.006 & $-$08:04:35.29 &                   &                  &                   &                   \\
 2 & 12:31:10.321 & $-$08:03:56.69 &   18.643  (0.010) &  17.836  (0.056) &  17.527   (0.114) & 17.329   (0.064)  \\
 3 & 12:31:15.903 & $-$08:02:14.07 &   16.916  (0.046) &  16.491  (0.054) &  16.332   (0.066) & 16.257   (0.047)  \\
 4 & 12:31:15.307 & $-$08:01:58.17 &   15.431  (0.035) &  14.905  (0.056) &  15.565   (2.424) & 14.687   (0.046)  \\
 5 & 12:31:08.100 & $-$08:02:01.29 &   16.593  (0.025) &  16.042  (0.041) &  15.858   (0.050) & 15.783   (0.048)  \\
 6 & 12:31:03.358 & $-$08:02:13.81 &   13.335  (0.027) &  12.666  (0.039) &  12.423   (0.045) & 12.295   (0.027)  \\
 7 & 12:30:57.996 & $-$08:01:32.58 &   15.960  (0.024) &  15.405  (0.035) &  15.200   (0.043) & 15.101   (0.036)  \\
 8 & 12:30:57.482 & $-$08:05:28.19 &   17.889  (0.088) &  16.399  (0.042) &  15.463   (0.053) & 14.992   (0.039)  \\
 9 & 12:31:01.452 & $-$08:06:14.02 &   14.700  (0.035) &  14.132  (0.031) &  13.985   (0.038) & 13.926   (0.037)  \\
10 & 12:31:07.750 & $-$08:06:39.12 &   16.723  (0.049) &  16.141  (0.046) &  15.994   (0.068) & 15.906   (0.061)  \\
11 & 12:31:19.407 & $-$08:07:03.51 &                   &                  &                   &                   \\
12 & 12:31:22.057 & $-$08:06:19.63 &   17.465  (0.109) &  16.097  (0.043) &  15.128   (0.054) & 14.650   (0.051)  \\
13 & 12:31:25.355 & $-$08:06:37.04 &                   &  15.677  (0.049) &  15.600   (0.115) & 15.505   (0.052)  \\
14 & 12:31:24.593 & $-$08:03:59.05 &   19.876  (0.010) &  13.723  (0.063) &  13.617   (0.053) & 13.590   (0.032)  \\
15 & 12:31:14.787 & $-$08:03:38.06 &   12.926  (0.026) &  12.426  (0.049) &  12.314   (0.045) & 12.270   (0.029)  \\
16 & 12:31:27.718 & $-$08:02:34.53 &   19.954  (0.010) &  11.940  (0.026) &  11.645   (0.059) & 11.456   (0.033)  \\
17 & 12:31:28.057 & $-$08:01:05.10 &                   &  16.027  (0.059) &  15.739   (0.050) &                   \\
18 & 12:31:21.260 & $-$08:00:22.14 &   17.396  (0.053) &  16.982  (0.025) &  16.894   (0.058) & 16.825   (0.057)  \\
19 & 12:31:21.166 & $-$08:00:11.70 &   17.893  (0.073) &  16.600  (0.039) &  15.808   (0.069) & 15.358   (0.028)  \\
20 & 12:31:17.638 & $-$07:59:52.25 &   16.181  (0.047) &  15.793  (0.053) &  15.710   (0.055) & 15.673   (0.058)  \\
21 & 12:31:12.089 & $-$08:00:13.41 &   15.088  (0.027) &  14.665  (0.066) &  14.579   (0.049) & 14.534   (0.039)  \\
22 & 12:31:06.949 & $-$07:59:24.83 &   16.754  (0.027) &  16.378  (0.036) &  16.280   (0.047) & 16.246   (0.050)  \\
23 & 12:31:05.295 & $-$07:59:05.75 &   17.425  (0.057) &  16.651  (0.042) &  16.375   (0.074) & 16.202   (0.065)  \\
24 & 12:31:05.463 & $-$07:58:54.88 &   17.900  (0.125) &  16.612  (0.038) &  16.050   (0.055) & 15.715   (0.058)  \\

\hline				  
\end{tabular}			  
\end{minipage}
\end{table*}

\begin{table*}
 \centering
 \begin{minipage}{170mm}
  \caption{$YJH$ magnitudes of local sequence stars.}
  \label{std_nir}
  \begin{tabular}{@{}cccccc@{}}
  \hline
  \hline
  Star & $\alpha_{\rm J2000}$ & $\delta_{\rm J2000}$ & $Y$ & $J$ & $H$  \\
 \hline
A & 12:33:29.431 & -08:19:01.37 &                & 13.687 (0.029) & 13.223 (0.022) \\
B & 12:33:46.507 & -08:14:55.17 &                & 10.875 (0.041) &                \\
C & 12:33:38.513 & -08:18:46.20 &                & 11.330 (0.029) & 10.834 (0.039) \\
D & 12:33:49.944 & -08:20:10.46 & 11.634 (0.023) & 11.467 (0.029) & 11.100 (0.019) \\
E & 12:33:59.758 & -08:20:31.31 & 13.017 (0.016) & 12.833 (0.013) & 12.534 (0.015) \\
F & 12:33:36.620 & -08:22:46.41 &                & 13.056 (0.036) & 12.710 (0.017) \\
G & 12:33:57.231 & -08:22:51.81 & 13.825 (0.023) & 13.395 (0.018) & 12.843 (0.017) \\
H & 12:33:47.233 & -08:16:46.01 & 13.909 (0.023) & 13.707 (0.017) & 13.301 (0.016) \\
I & 12:33:50.461 & -08:18:30.49 & 14.036 (0.016) & 13.794 (0.013) & 13.384 (0.012) \\
J & 12:33:32.643 & -08:22:00.73 &                & 13.751 (0.029) & 13.169 (0.022) \\
K & 12:33:51.115 & -08:24:45.67 & 14.422 (0.023) & 14.238 (0.019) & 13.932 (0.020) \\
L & 12:33:33.144 & -08:18:05.07 &                & 14.152 (0.030) & 13.687 (0.022) \\
M & 12:33:40.223 & -08:24:53.96 & 14.623 (0.023) & 14.226 (0.019) & 13.645 (0.019) \\
N & 12:33:56.318 & -08:16:44.15 & 14.556 (0.023) & 14.139 (0.017) & 13.501 (0.016) \\
O & 12:34:00.534 & -08:23:09.27 & 14.911 (0.023) & 14.679 (0.019) & 14.340 (0.023) \\
P & 12:33:52.784 & -08:16:24.71 & 15.066 (0.023) & 14.861 (0.018) & 14.465 (0.018) \\
Q & 12:33:42.916 & -08:23:11.40 & 15.272 (0.024) & 14.982 (0.019) & 14.574 (0.021) \\
R & 12:33:35.132 & -08:14:41.83 &                & 14.891 (0.039) & 14.271 (0.033) \\
S & 12:33:50.302 & -08:14:51.64 & 15.319 (0.024) & 15.014 (0.037) & 14.487 (0.020) \\
T & 12:33:30.343 & -08:19:45.39 &                & 15.306 (0.032) & 14.821 (0.028) \\
U & 12:33:56.389 & -08:14:46.54 & 15.557 (0.026) & 15.325 (0.021) & 14.976 (0.032) \\
V & 12:33:51.064 & -08:18:46.28 & 15.628 (0.018) & 15.378 (0.014) & 14.970 (0.017) \\
\hline				  
\end{tabular}			  
\end{minipage}
\end{table*}

\begin{table*}
 \centering
 \begin{minipage}{170mm}
  \caption{NIR photometry of SN 2009N.}
  \label{nirlc_table}
  \begin{tabular}{@{}ccccc@{}}
  \hline
  \hline
  date  & JD$-2400000$ & $Y$ & $J$ & $H$  \\
 \hline
26/01/2009 & 54857.8 & 15.738 (0.017) & 15.604 (0.014) & 15.369 (0.023) \\
29/01/2009 & 54860.8 & 15.709 (0.014) & 15.518 (0.018) & 15.278 (0.022) \\
24/02/2009 & 54886.8 & 15.306 (0.018) & 15.070 (0.017) & 14.751 (0.015) \\
25/02/2009 & 54887.8 & 15.291 (0.017) & 15.047 (0.017) & 14.747 (0.017) \\
28/02/2009 & 54890.8 & 15.268 (0.015) & 15.007 (0.014) & 14.739 (0.017) \\
05/03/2009 & 54895.9 & 15.206 (0.014) & 14.971 (0.018) & 14.649 (0.019) \\
11/03/2009 & 54901.7 & 15.180 (0.014) & 14.941 (0.017) & 14.620 (0.015) \\
22/03/2009 & 54912.9 & 15.197 (0.014) & 14.927 (0.018) & 14.590 (0.016) \\
29/03/2009 & 54919.8 & 15.213 (0.023) & 14.919 (0.018) & 14.600 (0.021) \\
24/04/2009 & 54945.7 & 15.558 (0.014) & 15.298 (0.015) & 14.910 (0.017) \\
19/03/2010 & 55274.7 & 20.538 (0.378) & 20.171 (0.239) & 19.110 (0.390) \\
19/03/2010 & 55274.8 & 20.602 (0.335) & 20.540 (0.487) & \\
\hline				  
\end{tabular}			  
\end{minipage}
\end{table*}

\begin{table*}
 \centering
 \begin{minipage}{170mm}
  \caption{{\it Swift} photometry of SN 2009N}
  \label{swift_lc}
  \begin{tabular}{@{}cccccccc@{}}
  \hline
  \hline
date & JD$-2400000$ & $uvw2$ & $uvm2$ & $uvw1$ & $u$ & $b$ & $v$\\
\hline
27/01/2009 & 54858.8 &  16.885 (0.068) & 16.711 (0.062) & 16.200 (0.053) & 15.599 (0.047) & 16.541 (0.043) & 16.351 (0.049) \\
29/01/2009 & 54861.0 &  17.618 (0.099) & 17.615 (0.137) & 16.848 (0.082) & 15.873 (0.059) & 16.657 (0.057) & 16.423 (0.077) \\
01/02/2009 & 54863.8 &  18.404 (0.089) & 18.477 (0.121) & 17.664 (0.072) & 16.254 (0.049) & 16.693 (0.043) & 16.298 (0.048) \\
04/02/2009 & 54867.2 &  18.740 (0.100) & 18.792 (0.130) & 18.047 (0.082) & 16.892 (0.055) & 16.820 (0.044) & 16.394 (0.048) \\
16/03/2009 & 54907.3 &  19.035 (0.107) & 18.996 (0.146) & 19.069 (0.134) & 18.781 (0.113) & 17.765 (0.051) & 16.474 (0.046) \\

\hline				  
\end{tabular}			  
\end{minipage}
\end{table*}

\end{document}